\documentclass[aps,prd,twocolumn,floatfix,nofootinbib,superscriptaddress,tightenlines]{revtex4}
\usepackage[dvipsnames]{xcolor}
\usepackage{amsmath}
\usepackage{dcolumn}
\usepackage{lipsum}
\usepackage{amssymb}
\usepackage{url}
\usepackage{epsfig}
\usepackage{graphicx}
\usepackage{amsmath}
\usepackage{bm}
\usepackage{setspace}
\usepackage{lscape}
\usepackage{amsthm}
\usepackage{bbold}
\usepackage{dcolumn}
\usepackage{epsfig}
\usepackage{graphics}
\usepackage{graphicx}
\usepackage{longtable}
\usepackage{svg}

\usepackage{bm}
\usepackage{xspace}
\usepackage{cancel}
\usepackage{float}
\usepackage{multirow}
\definecolor{darkgreen}{rgb}{0,0.5,0}
\definecolor{purple}{rgb}{0.5,0,0.5}
\definecolor{nblue}{rgb}{0.0,0.0,0.50}
\definecolor{scarlet}{rgb}{1.0,0.2,0}
\definecolor{darkmagenta}{rgb}{0.55, 0.0, 0.55}
\definecolor{darkolivegreen}{rgb}{0.33, 0.42, 0.18}
\definecolor{darkcandyapplered}{rgb}{0.64, 0.0, 0.0}
\usepackage[colorlinks=true, pdfstartview=FitV, linkcolor=purple, citecolor= purple, urlcolor=blue]{hyperref}

\def\beq{\begin{equation}} \def\eeq{\end{equation}}
\def\beqn{\begin{eqnarray}} \def\eeqn{\end{eqnarray}}

\def\beq{\begin{equation}} 
\def\eeq{\end{equation}} 
\def\beqn{\begin{eqnarray}} 
\def\eeqn{\end{eqnarray}} 
 
\def\to{\rightarrow}
\def\nn{\nonumber}

\begin{document}
\title{Phenomenological extraction of fragmentation functions in a $p\bar{p}$ environment}

\author{M.~A.~Pérez~de~León}
\email{marioaldair.perez@uas.edu.mx}
\affiliation{Facultad de Ciencias F\'isico-Matem\'aticas, Universidad Aut\'onoma de Sinaloa, Ciudad Universitaria, Culiac\'an, Sinaloa 80000,
M\'exico.}
\affiliation{Facultad de Ciencias de la Tierra y el Espacio, Universidad Aut\'onoma de Sinaloa, Ciudad Universitaria, Culiac\'an, Sinaloa 80000,
M\'exico.}

\author{S.~A.~Ochoa-Oregon}
\email{salvadorochoa.fcfm@ms.uas.edu.mx}
\affiliation{Facultad de Ciencias F\'isico-Matem\'aticas, Universidad Aut\'onoma de Sinaloa, Ciudad Universitaria, Culiac\'an, Sinaloa 80000,
M\'exico.}
%\affiliation{Instituto de Física Corpuscular, Universitat de Valencia - Consejo Superior de Investigaciones Científicas,
%Parc Científic, E-46980 Paterna, Valencia, Spain} 

\author{D.~F.~Renter\'ia-Estrada}
\email{david.renteria@ific.uv.es}
\affiliation{Instituto de Física Corpuscular, Universitat de Valencia - Consejo Superior de Investigaciones Científicas,
Parc Científic, E-46980 Paterna, Valencia, Spain} 

\author{R.~J.~Hern\'andez-Pinto}
\email{roger@uas.edu.mx}
\affiliation{Facultad de Ciencias F\'isico-Matem\'aticas, Universidad Aut\'onoma de Sinaloa, Ciudad Universitaria, Culiac\'an, Sinaloa 80000,
M\'exico.}

%\author{German~F.~R. Sborlini}
%\email{german.sborlini@usal.es}
%\affiliation{Departamento de Física Fundamental e IUFFyM, Universidad de Salamanca, Plaza de la Merced S/N, 37008 Salamanca, Spain.}

\begin{abstract}
The precise determination of fragmentation functions (FFs) of hadrons relies on the accurate description of the differential cross sections obtained from both experimental high-energy hadron colliders and theoretical predictions at higher orders in quantum chromodynamics. Various phenomenological strategies have been employed to extract FFs. In this work, we analyze the use of kinematical cuts for reactions including pions and kaons in proton-antiproton collisions to isolate individual FF contributions. This study examines the feasibility of using a similar approach as in proton-proton colliders to analyze FF flavour separation~\cite{Ochoa-Oregon:2023ktx}. In particular, we study photon-hadron production at colliders, including NLO QCD and LO QED corrections to reconstruct the partonic momentum fractions. 
\end{abstract}

\maketitle               

%\tableofcontents

%%%%%%%%%%%%%%%%%%%%%%%%%%%%%%%%%%%%%%%%%%
%%%%%%%%%%%%%%%%%%%%%%%%%%%%%%%%%%%%%%%%%%
\section{Introduction and motivation}
\label{sec:introduction}
One of the great challenges currently facing the scientific community is understanding the elementary particles that make up everything we know. Through theoretical and experimental collaborations, high-energy physicists have established that the Standard Model (SM) of elementary particles is an incredibly precise description of the fundamental building blocks of matter; nonetheless, some aspects still escape to it. 

SM is a quantum theory of non-abelian fields that describes the dynamic properties of particles and their interactions~\cite{Weinberg:1967tq}. This theory lacks exact solutions, so it is key to find mechanisms that can provide highly accurate theoretical predictions to compare with experimental data. Perturbative theoretical expansions allow us to reach the experimental accuracy of measurements asymptotically and thus verify the validity of the SM. The factorization theorem of Quantum Chromodynamics (QCD) has successfully guided major discoveries such as the identification of the Higgs boson~\cite{ATLAS:2012yve,CMS:2012qbp}. This discovery remains of interest when it comes to testing the validity of the SM or finding theories beyond the SM (BSM). According to the factorization formula, physical observables in high-energy hadron colliders can be described by the convolution of partonic cross sections, calculated in pQCD, and distribution functions obtained from global QCD analyses~\cite{Collins:1989gx}. Although calculating high-order corrections to partonic cross sections is a key area of research in theoretical high-energy physics, this document aims to deepen our understanding of non-perturbative QCD distribution functions.

%The factorization theorem of Quantum Chromodynamics (QCD) has successfully guided major discoveries, such as the discovery of the Higgs boson, among others~\cite{ATLAS:2012yve,CMS:2012qbp}. This discovery remains a subject of interest to test the validity of the SM or to find theories beyond the SM (BSM). The factorization formula states that physical observables in high-energy hadron colliders can be described by the convolution of partonic cross sections, calculated in pQCD, and probability distribution functions, obtained from global QCD analyses~\cite{Collins:1989gx}. Although the calculation of high order corrections to partonic cross sections is one of the main active research areas in theoretical High-Energy Physics, in this document we are interested in deepening our understanding of the non-perturbative QCD distribution functions.

In particle physics phenomenology, long-distance (non-perturbative)
dynamics are encoded in distribution functions which are classified into two categories: {\it i)} parton distribution functions (PDFs), $f_a^h(x,\mu)$, which represent the probability of extracting a parton $a$ from a hadron $h$ with momentum fraction $x$ at the scale $\mu$, and {\it ii)} fragmentation functions (FFs), $d_a^h(z,\mu)$, which characterize the probability of producing a hadron $h$ with momentum fraction $z$ from the hadronization of a parton $a$ at the scale $\mu$. 
Their determination with high precision is a challenging task and crucial for the understanding of high-energy collider phenomena. Since these functions cannot be fully derived from first principles, their extraction requires the combined use of theoretical predictions and experimental measurements. From the experimental side, data sets are extracted from various processes, such as: {Drell-Yan production (DY) \cite{Moreno:1990sf,NuSea:2003qoe,NuSea:2001idv,CDF:2010vek,SeaQuest:2021zxb}, deep inelastic scattering (DIS) \cite{NewMuon:1996uwk,NewMuon:1996fwh,Whitlow:1991uw,BCDMS:1989qop,CHORUS:2005cpn,NuTeV:2001dfo},} semi-inclusive deep inelastic scattering (SIDIS) \cite{HERMES:2012uyd,COMPASS:2016xvm,COMPASS:2016crr} and hadron–hadron collisions \cite{STAR:2013zyt,STAR:2011iap,STAR:2009vxb,STAR:2006xud,ALICE:2012wos,ALICE:2014juv,PHENIX:2007kqm} in the case of PDFs, while single-inclusive annihilation (SIA) \cite{OPAL:1994zan,TASSO:1988jma,BaBar:2013yrg,Belle:2013lfg,ALEPH:1994cbg,DELPHI:1998cgx,SLD:1998coh,OPAL:1999zfe,TPCTwoGamma:1986evp,TPCTwoGamma:1988yjh} is also relevant to generate global QCD analyses of FFs. On the other hand, theoretical calculations of partonic cross sections for the three processes are computed at the highest precision in pQCD in order to connect, through the factorization formula, partonic with hadronic observables \cite{Hirai:2007cx,Albino:2005mv,Albino:2005me,Kniehl:2000fe,Kretzer:2000yf,deFlorian:2014xna,deFlorian:2017lwf,Borsa:2022vvp,AbdulKhalek:2022laj,Ritzmann:2014mka}. The consistency between theory and experiment ultimately relies on the specific modelling of PDFs and FFs. Parametrizations of distribution functions were mainly based on phenomenological ansatzes; nowadays, since computational tools are evolving extremely fast, recent extractions of distribution functions implement searches for best fitting parameters through neural networks where no ansatz is needed~\cite{Metz:2016swz,NNPDF:2021njg}. 

Global QCD fits are relevant since they test one of the key features of all the distribution functions, universality. Although most of the groups achieve reasonable agreement with the experimental data within the corresponding uncertainty bands, the question of proper theoretical description of PDFs and FFs remains open. Moreover, it is well established that incorporating measurements at different energy scales (e.g., RHIC at 500 GeV and LHC at 13 TeV) may introduce tensions between models, challenging the naive universality expected for PDFs and FFs. When the processes considered in the theory–experiment comparison are sufficiently inclusive, such differences can be mitigated through adjustments of the factorization scales \cite{Borsa:2021ran}. However, when less inclusive observables are involved, the problem becomes more severe, setting a significant challenge for the understanding of hadronization and motivating the development of novel approaches to describe this phenomenon.
In light of this context, the present work aims to investigate strategies that allow for tighter constraints on FFs. Our study builds on the work in the Ref.~\cite{Ochoa-Oregon:2023ktx}, which clearly establishes a methodology for determining flavored parton fragmentation functions through kinematic cuts in final-state particles. Ref.~\cite{Ochoa-Oregon:2023ktx} analyzes proton collisions at LHC and FCC energies and in this study, we apply the referenced methodology to proton-antiproton collisions at Tevatron energies. 

From a theoretical perspective, all partonic channels are relevant to the hadronic cross-section in a proton-proton collider operating at very high energies. In a proton-antiproton collider with a centre-of-mass energy of 1.986 TeV, however, valence quarks make the most significant contribution to the observables and the initial state of the beams is asymmetric. These two characteristics make studying collisions in the Tevatron configuration crucial for understanding the mechanisms that differentiate the fragmentation functions of hadrons originating from valence quarks. Despite the energy being an order of magnitude lower {with respect to the LHC}, the Tevatron environment has the potential to increase the sensitivity of fragmentation functions to different kinematic cuts. The main motivation of this paper is to find relations between parton-to-pion and parton-to-kaon FFs from the study of ratios of the $z$-spectra of $p+\bar{p} \to \gamma + \pi$ and $p+\bar{p} \to \gamma + K$ when kinematical cuts are applied.

The outline of this paper is as follows: Sec.~\ref{sec:Analysis} provides a detailed description of the Monte Carlo simulations implemented.
It also explains the impact of different cuts in cross section distributions to distinguish features on each production channel. Sec.~\ref{sec:FFconstraints} compares the ratio of cross section distributions in the reconstructed momentum fraction, $z$, with respect to FF ratios. We discuss the similarities among these plots and suggest ways to improve the FF constraint from cross-section shapes. In particular, Sec.~\ref{ssec:FFimproved} uses the reconstructed momentum fractions discussed in Refs.~\cite{deFlorian:2010vy,Renteria-Estrada:2022scipost} to impose specific cuts and extract information about FF ratios {; subsequently, Section~\ref{ssec:IIIC} extends this analysis by incorporating the next subleading term into the approximation.} Finally, Sec.~\ref{sec:conclusions} presents the conclusions of this work and proposes strategies to improve the quality of kaon and other heavier hadron fragmentation functions for future research programs.

%%%%%%%%%%%%%%%%%%%%%%%
%%%%%%%%%%%%%%%%%%%%%%%
%%%%%%%%%%%%%%%%%%%%%%%
\section{Computational details and phenomenological analysis}
\label{sec:Analysis}

Our starting point is the factorization formula. Consider the collision of two hadrons that generates a third hadron in association with a hard photon. Thus, the reaction is,
\begin{align}
    H_1(P_1)+H_2(P_2)\to h(P_h)+\gamma(P_\gamma) {+X}
\end{align}
where $P_{1,2}$ are the four momenta of the incoming hadrons and $P_h$ and $P_\gamma$ are the four momenta of the hadron $h$ and the photon in the final state, respectively.
According to the factorization theorem in pQCD, this process can be computed as
\begin{align}
    d\sigma_{H_1H_2\rightarrow h\gamma}=&\sum_{a,b,c}f_{a}^{H_1} \otimes f_{b}^{H_2} \otimes d\hat{\sigma}_{ab\rightarrow c\gamma}\otimes d_{c}^{h},
\end{align}
where the sum is over all contributing partonic channels $a + b \to c + \gamma$ , with $d\hat{\sigma}_{ab\rightarrow c\gamma}$ the associated partonic cross section. $d\hat{\sigma}_{ab\rightarrow c\gamma}$ can be expanded as a power series in the coupling constants of the SM. In this work, we are interested in Monte Carlo (MC) analyses with the partonic cross-section computed at first order pQCD, so called Next-to-Leading Order (NLO) in QCD, plus the zeroth order correction in Quantum Electrodynamics (QED), so called Leading Order (LO) in QED. 

In particular, we focus on the phenomenological determination of flavored parton-to-pion and parton-to-kaon FFs in an asymmetric environment, the proton-antiproton collider. We studied the collisions of protons and anti-protons at Tevatron center of mass energy of $\sqrt{s}=1.986$ TeV. Thus, we center our analysis in two clear processes,
\begin{align}
    p\,+\,\bar{p}&\rightarrow \gamma\,+ \pi, \nn \\
    p\,+\,\bar{p}&\rightarrow \gamma\,+ K,
\end{align}
where $\pi=\{\pi^\pm,\pi^0\}$ and $K=\{K^\pm,K^0\}$. Reconstruction of physical observables was achieved by the smooth-cone isolation algorithm \cite{Frixione:1998jh}. {This isolation mechanism defines an infrared-safe (IR-divergence-free) cut-off function. The cut function, defined as,}
\begin{align}
    \xi(r) = \epsilon_r\,E_T^\gamma\, \left( \frac{1-\cos r}{1-\cos r_0}\right)^4 \, ,
\end{align}
takes into account the amount of energy that the prompt photon is carrying. Furthermore, this cut-function takes into consideration the transverse energy of the photon $E_T^\gamma$ and two phenomenological parameters $\epsilon_r$ and $r_0$, which we set to 1.0 and 0.4, respectively, inspired by previous results~\cite{Ochoa-Oregon:2023ktx}. {Mathematically, this function replaces sharp boundaries, such as Heaviside step function, with a smooth distribution parameterized by the isolation distance $r$. 
This distance is defined in the pseudorapidity($\eta$)-azimuthal angle($\phi$) plane as,
\begin{align}
r=\sqrt{(\eta^h-\eta^\gamma)^2+(\phi^h-\phi^\gamma)^2} \,,
\end{align}
which sets the isolation distance between the final-state hadron and the hard photon.}
%%%%%%%%%%%%%%%%%%%%%%%%
%%%%%%%%%%%%%%%%%%%%%%%%
%%%%%%%%%%%%%%%%%%%%%%%%
{To present phenomenological results, it is necessary to define the procedure for estimating theoretical uncertainties in our numerical predictions. For a $2 \to 2$ process, it is standard practice to define the hard scale of the process as the average transverse energy of the two final-state particles. From a theoretical perspective, however, such a choice causes the evolution of the PDFs, via the DGLAP equations \cite{Gribov:1972ri,Altarelli:1977zs,Dokshitzer:1977sg}, to exhibit an undesirable dependence on the momentum fraction of the hadronized parton, $z$. To address this concern, we have chosen the scale $\mu = p_T^\gamma$, thereby avoiding a direct dependence on $z$.}
%The energy scale of each process is given by the average transverse momentum of the hadron and the photon, i.e.
%\begin{equation}
%\label{mu}
%    \mu=\frac{p_T^\gamma+p_T^h}{2} \, \, ,
%\end{equation}
%with $h=\pi, K$. 
Then, in our simulations we set the scales as $\mu=\mu_R=\mu_I=\mu_F$, where {$\mu_R$} stands for the renormalization scale, while $\mu_I$ and final $\mu_F$ for the initial and final factorization scales.
{It should be noted that the theoretical uncertainties considered in this work are restricted to those arising from the variation of the hard scales. While the impact of uncertainties stemming from the parton distribution functions (PDFs and FFs) is significant, its detailed evaluation is deferred to future work, as the current methodology is sufficient to substantiate the findings presented in this work.}

In addition to the Tevatron center of mass energy, $E_{C.M.}=\sqrt{S_{C.M.}}=1.986\, \text{TeV}$, we also consider the following kinematical cuts,
\begin{align}\label{eq:cuts}
    \{|\eta^h|,|\eta^\gamma|\}\,&\leq\,2.5, \nn\\
    2\,\text{GeV}\leq p_T^h &\leq \, 15\,\text{GeV},\\
    p_T^\gamma &\geq\,30\,\text{GeV}.\nn
\end{align}
PDFs were implemented through the LHAPDF framework; in particular for our simulations we have used the PDF sets: \texttt{NNPDF40\_nlo\_as\_01180} \cite{Kassabov:2022pps} for LO QCD and, whereas for LO QED + NLO QCD corrections, we used \texttt{NNPDF31\_nlo\_as\_0118\_luxqed}~\cite{Bertone:2017bme,Campbell:2018wfu,Manohar:2017eqh}. The \texttt{DSS2014}~\cite{deFlorian:2014xna} and \texttt{DSS2017}~\cite{deFlorian:2017lwf} routines were used for the hadronization process for pions and kaons, respectively.
%%%%%%%%%%%%%%%%%
%%%%%%%%%%%%%%%%%
%%%%%%%%%%%%%%%%%
\subsection{Differential cross-section as a function of the partonic momentum fractions}
\label{ssec:DistributionsZ}

Our main objective is the extraction of flavored hadron FFs by analyzing cross section distributions obtained from MC simulations. Specifically, we are interested in pions and kaons because since they are the cleanest signature of identified hadrons in detectors. Our initial step consists in analyzing the $z$-spectrum of 
hadronic cross sections. Since our study is based on theoretical MC simulations, we are able to know the momentum fraction carried by the hadron, $z_{\rm REAL}$, with certainty. However, to develop a methodology based on experimental observables, we will use a reconstructed momentum fraction, $z_{\rm REC}$, which is a function of the particles' momenta in the identified final state. {Guided by the kinematic structure at LO}, we define,
\begin{equation}
    z_\text{REC}=\frac{p^h_T}{p^\gamma_T} \, ,
\end{equation}
since it allows for simpler implementation and it is closely related with with alternative approximations \cite{deFlorian:2010vy,Renteria-Estrada:2021rqp}. {While this definition is strictly valid at LO, it is expected to break down upon the inclusion of higher order corrections. A primary objective of this work is to identify regions where the reconstructed functions remain applicable, enabling the extraction of relevant information regarding parton kinematics.} In addition, it is important to note that we restrict our analysis to the region $z \in (0.1,0.8)$ since it is the region where observables are better understood phenomenologically.

%\bl{This choice is made because it enables a simpler implementation, while yielding results that are in close agreement with alternative approximations reported in Refs. [49,51]. Our study is limited to the region $z \in (0.1,0.8)$ for two primary reasons. First, the cross section becomes strongly suppressed outside this interval as a consequence of kinematical constraints. Second, the FFs are considered reliable only within this domain, and any extrapolation beyond it could result in unphysical interpretations.}

%\bl{Before proceeding with the analysis, we emphasize the motivation for comparing the $z$-spectrum of $\gamma+\pi^{\pm}$ with $\gamma+K^{\pm}$, as well as $\gamma+\pi^{0}$ with $\gamma+K^{0}$. Since the objective of this work is to establish connections between pion and kaon FFs, it is natural to expect similarities between $K^{\pm}$ and $\pi^{\pm}$, given that they share identical electromagnetic and isospin charges. Nonetheless, while the set $\{\pi^+,\pi^-,\pi^0\}$ forms an isospin multiplet, the same does not hold for $\{K^+,K^-\}$ and $K^0$. In contrast, both $K^0$ and $\pi^0$ are neutral states, which suggests that possible electromagnetic effects may be reduced—or even canceled—when comparing $\gamma+\pi^0$ and $\gamma+K^0$. As will be clarified later in this article, our methodology is ultimately independent of the specific nature of the hadrons being compared.}
%%%%
%%%con deppl
%%%

Before proceeding with the analysis, we would like to emphasise the reasons behind comparing the $z$-spectrum of $\gamma+\pi^\pm$ with that of $\gamma+K^\pm$, as well as $\gamma+\pi^0$ with $\gamma+K^0$. As this study aims to identify links between fragmentation functions of pions and kaons, it is reasonable to anticipate similarities between $K^+$ and $K^-$, given that they have the same electromagnetic {coupling} and isospin. However, while the set $\{\pi^+,\pi^-,\pi^0\}$ forms an isospin multiplet, this is not the case for the kaon system. In contrast, both $K^0$ and $\pi^0$ are neutral states, suggesting that possible electromagnetic effects may be reduced or even cancelled out  when comparing $\gamma+\pi^0$ and $\gamma+K^0$. As will be explained later in this article, the theoretical nature of the compared hadrons does not constraint the way we implement our research; thus, we rely purely on the phenomenological description of observables in High-Energy collider experiments to proceed with the extraction of FFs.
\begin{figure*}[t!]
    \centering
    \includegraphics[width=0.49\textwidth]{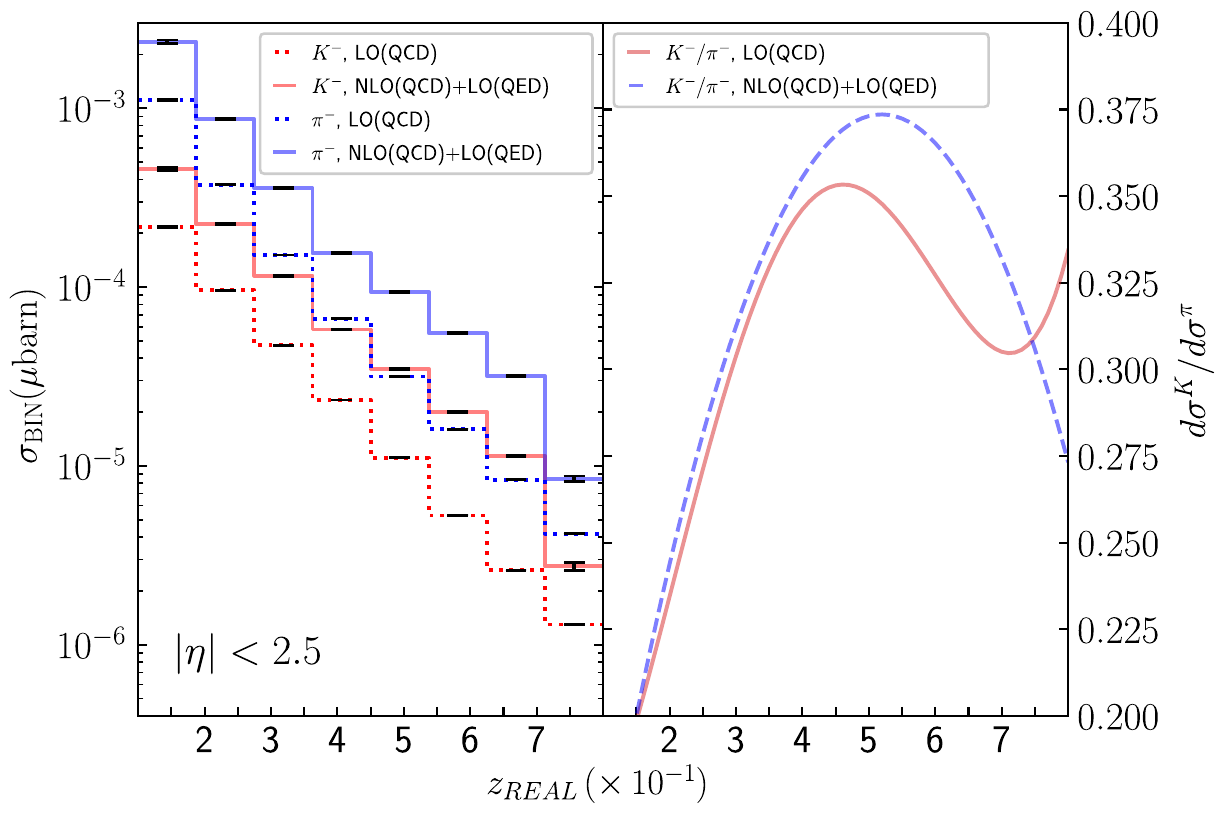}
    \includegraphics[width=0.49\textwidth]{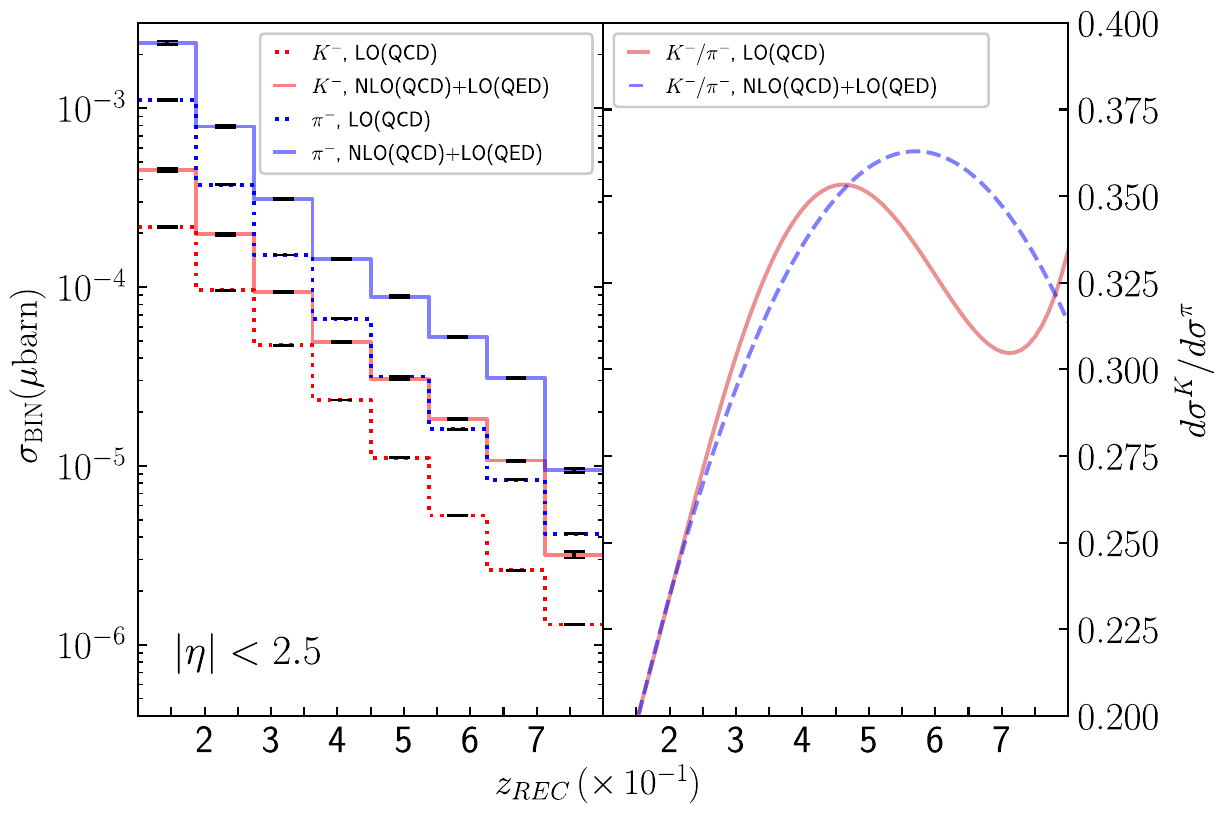}
    \includegraphics[width=0.49\textwidth]{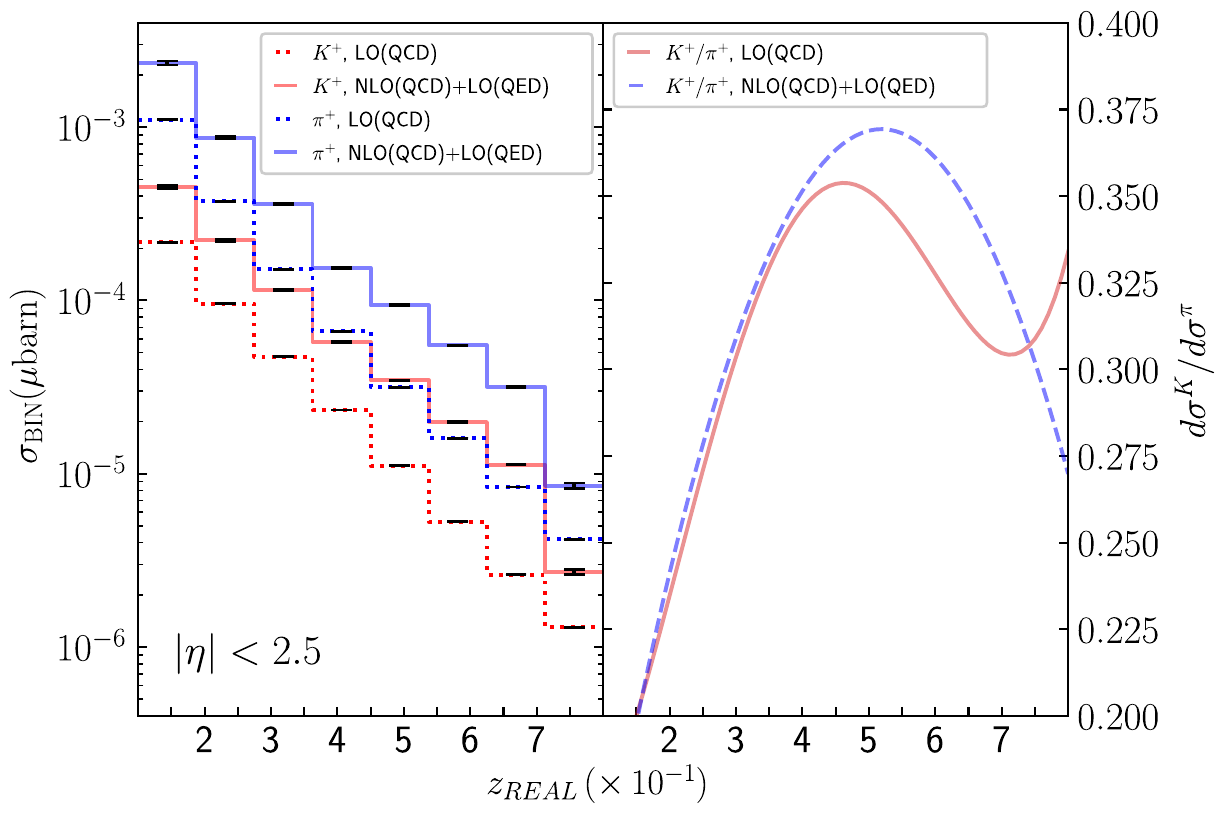}
    \includegraphics[width=0.49\textwidth]{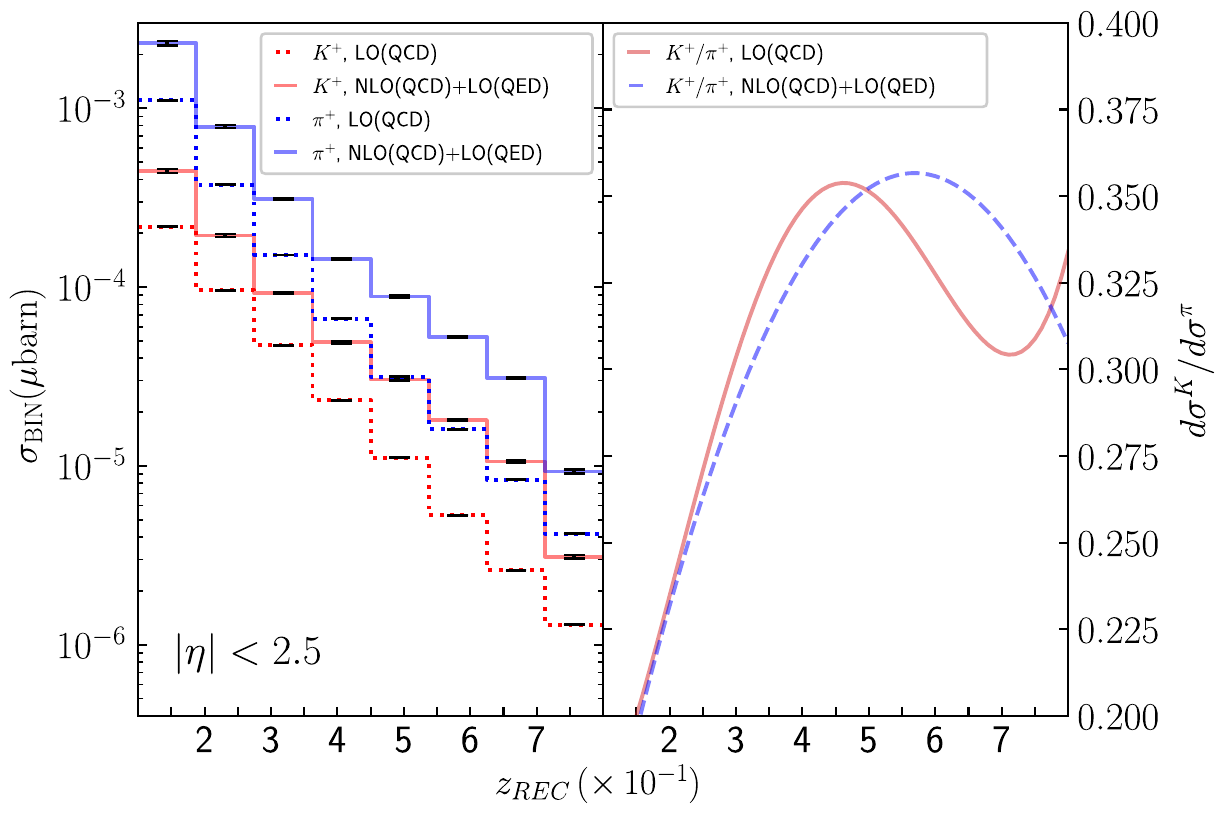}
    \includegraphics[width=0.49\textwidth]{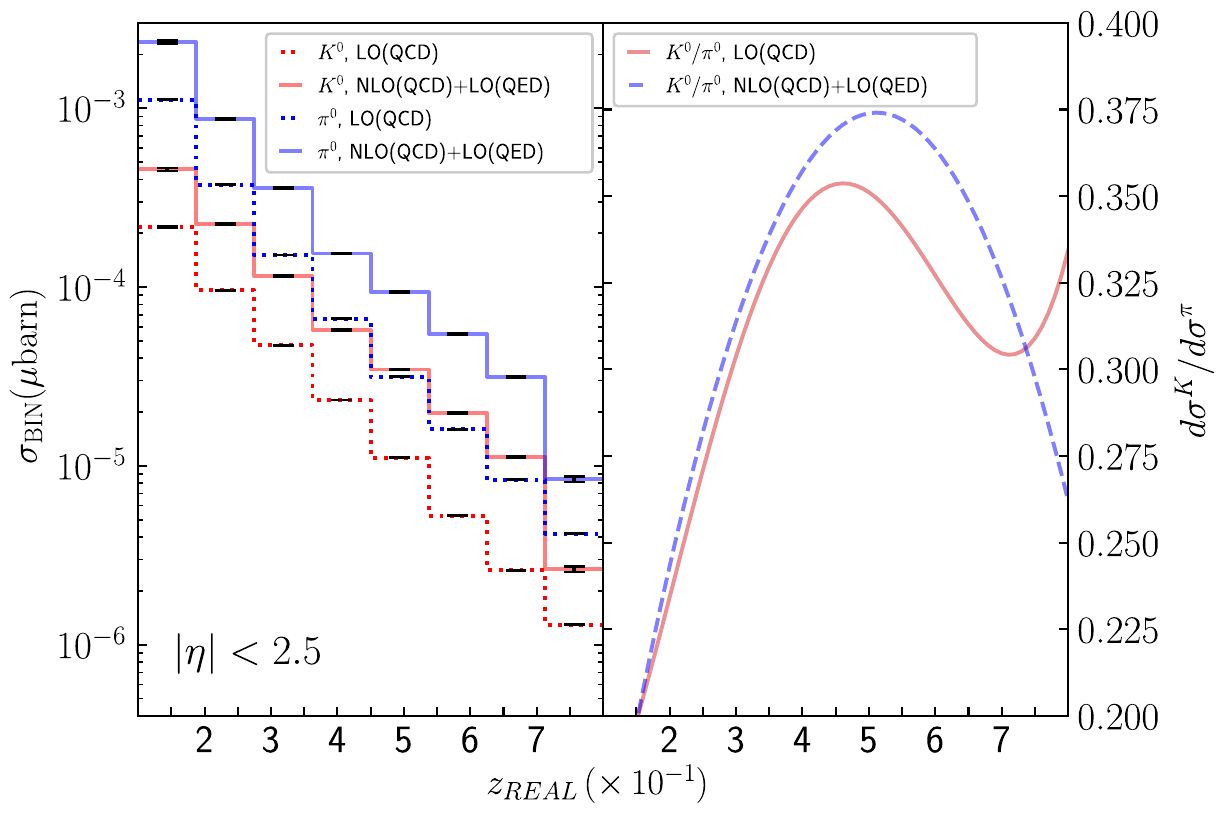}
    \includegraphics[width=0.49\textwidth]{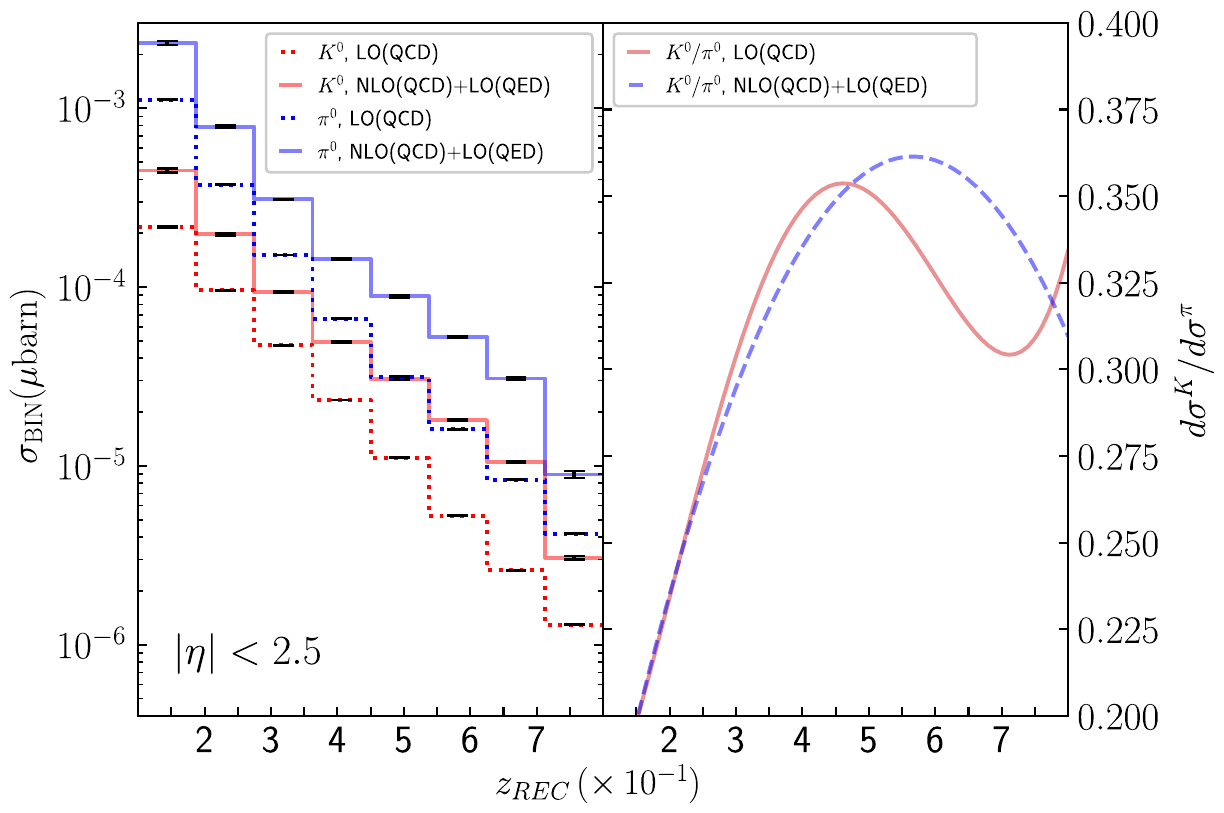}
    \caption{Differential cross-section distribution of pions and kaons as a function of $z_{REAL}$ ({left} column) and $z_{REC}$ ({right} column), using Tevatron kinematics ($\sqrt{S_\text{C.M.}} = 1.986 \, {\rm TeV}$). On the l.h.s. of each panel, we display the differential cross-sections, whilst on the r.h.s of each panel we present the ratio $d\sigma^K/d\sigma^\pi$ at LO(QCD) including NLO(QCD)+LO(QED) corrections. We present the results for negative (first row), positive (second row) and neutral (third row) hadron production. }
    \label{fig:Figura1}
\end{figure*}

%%%%%%%%%%%%%%
%%%%%%%%%%%%%%
%%%%%%%%%%%%%%
In Fig.~\ref{fig:Figura1} we present the results of the MC simulations of proton-antiproton collisions producing a photon in association with: {\it i)} negative (top), {\it ii)} positive (middle), and {\it iii)} neutral (bottom) pion and kaons as a function of $z_{\rm{REAL}}$ (l.h.s.) and $z_{\rm{REC}}$ (r.h.s.). Each panel shows on the {l.h.s.} differential cross-section at fixed order, LO QCD and NLO QCD + LO QED corrections, and, on the {r.h.s.} of each panel the ratio of differential cross-sections between kaons and pions. We recall that Fig.~\ref{fig:Figura1} shows the results when the kinematic cuts expressed in Eq.~(\ref{eq:cuts}) are considered. We highlight that differential cross-section distributions of initiated proton-antiproton collisions are one order of magnitude smaller than to proton-proton, see Ref. \cite{Ochoa-Oregon:2023ktx}. Since we are interested in the phenomenological description of differential cross-sections as a function of $z$, we point out that in general, LO (QCD) + NLO (QCD) ratios between a pions and kaons are in close agreement among $z_{\rm{REAL}}$ and $z_{\rm{REC}}$. Then, we can conclude from these first results that $z_{\rm{REC}}$ is a good phenomenological variable to perform QCD studies, when precision observables are required. In the following results we present our discussions in terms of $z_{\rm REC}$\footnote{Nonetheless, we shall present, when possible, a cross validation with $z_{\rm REAL}$.}. In addition to the $p_T$-cuts presented in Eq.~(\ref{eq:cuts}), we shall test the impact on the differential cross sections in: $\vert\eta\vert  < 0.5$ and {$1.5 < \vert\eta\vert  < 2$}. For further comparison, we name the case $\vert\eta\vert  < 0.5$ by {\it Scenario 1} while, {$1.5<\vert\eta\vert  < 2$} by {\it Scenario 2}.
Then, we examine the separation of signals by the study of pseudorapity cuts of {\it Scenario 2}. Before moving towards that direction, we present in the following subsection the analysis of different channels to the total cross-sections.

%%%%%%%%%%%%%%%%%%%%%%%%%%%%%%%%%%%%%%%%%%
%%%%%%%%%%%%%%%%%%%%%%%%%%%%%%%%%%%%%%%%%%
\subsection{Analysis of different parton-initiated contributions}
\label{ssec:DistributionsZparton}
The second part of the study analyses the influence of different parton contributions on hadron-photon production. The principal objective is the identification of the dominant channel and assesses how sensitive are the distributions to certain kinematic cuts.
\begin{figure}[t!]
    \centering
    \includegraphics[width=0.49\textwidth]{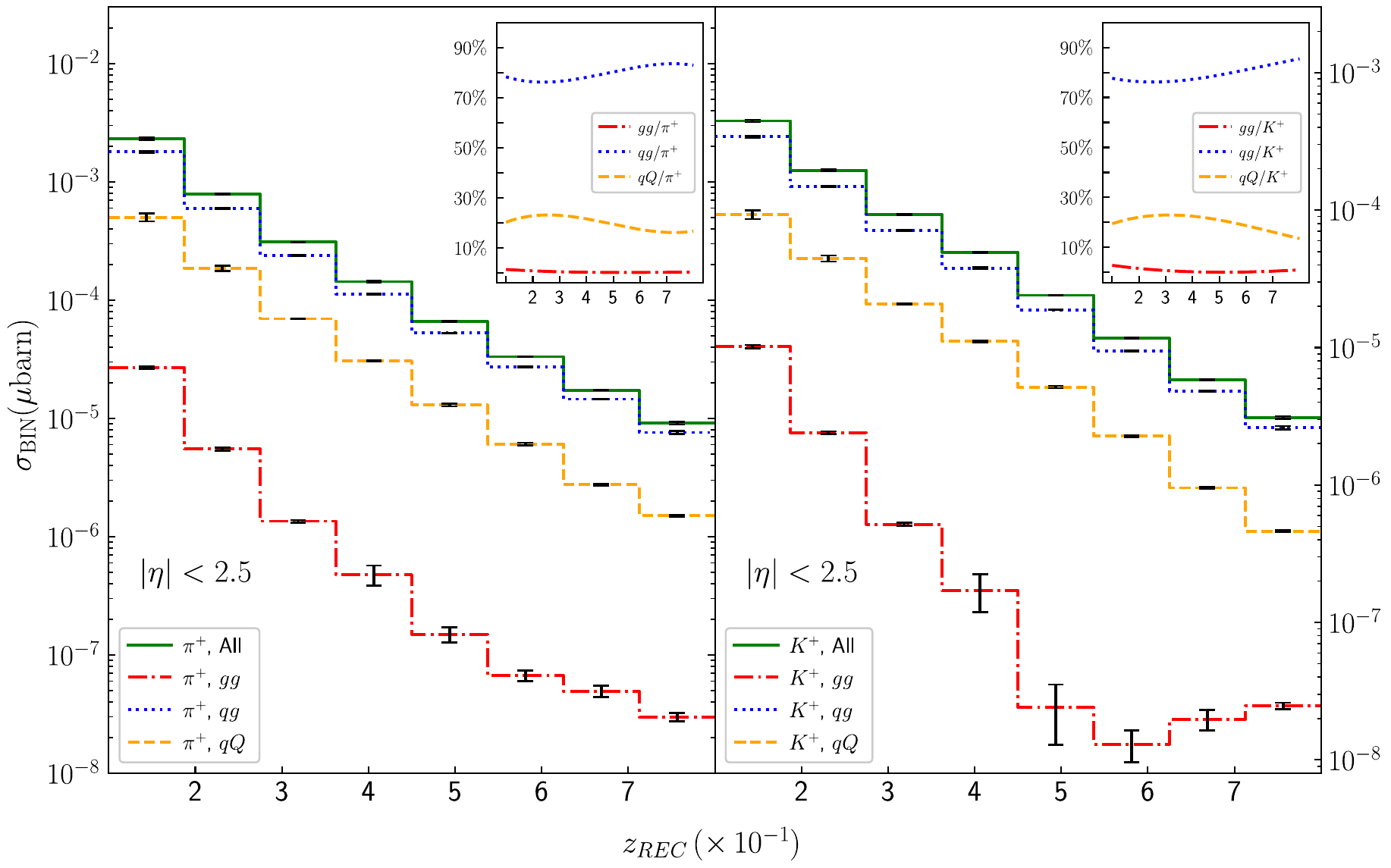}
    \vspace{0.5cm} % espacio vertical entre las figuras
    \includegraphics[width=0.49\textwidth]{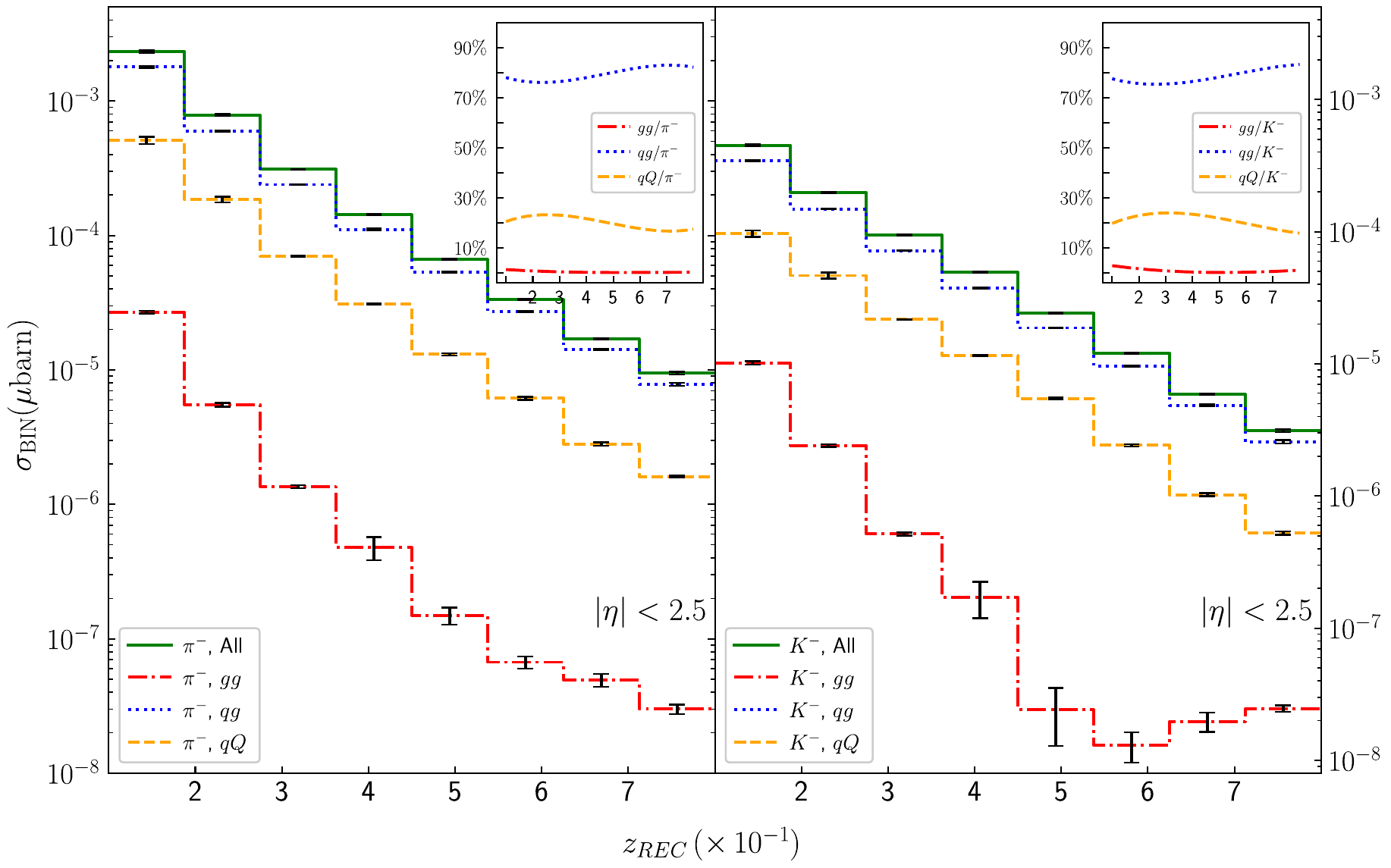}
    
    \caption{Differential cross-section distribution as a function of $z_\text{REC}$ for positive (top) and negative (bottom) charged hadrons production in the scenario defined by Eq.~(\ref{eq:cuts}). We indicate the contributions due to the different partonic channels: $gg$ (dashed red line), $qg$ (point blue line), and $qQ$ (dashed orange line). We use Tevatron kinematics ($\sqrt{S_\text{C.M.}}=1.938\,\text{TeV}$). In the upper sector of each plot, we indicate the relative contribution of each channel.}
    \label{fig:Figura2}
\end{figure}

We present in Fig.~\ref{fig:Figura2} the contribution of different initiated partonic channels to the total cross-section as a function of $z_{\rm REC}$. All calculations were performed with NLO (QCD) + LO (QED) corrections to $p+\bar{p}\to \gamma + h$, $h=\{\pi^{\pm},K^{\pm}\}$ with kinematic cuts expressed in Eq.~(\ref{eq:cuts}). In the upper row the production of a photon in association with positive charged hadrons is presented while, the photon production with negative charged hadrons is in the lower row. {We classify the production processes into three main categories based on their initial-state partons: the $qg$ channel (quark-gluon scattering), the $qQ$ channel (encompassing all quark-quark and quark-antiquark flavor combinations), and the $gg$ channel (gluon-gluon fusion). Our analysis shows that the $qg$ channel provides the dominant contribution, followed by the $qQ$ combinations, with the $gg$ process being the least significant.} %We can distinguish that the dominant channel is $qg$-channel, followed by $qQ$ and finishing with the $gg$-channel. 
These distributions will serve as a reference for comparing the sensitivity per channel.
\label{ssec:Subchannels}
\begin{figure}[t!]
    \centering
    \includegraphics[width=0.49\textwidth]{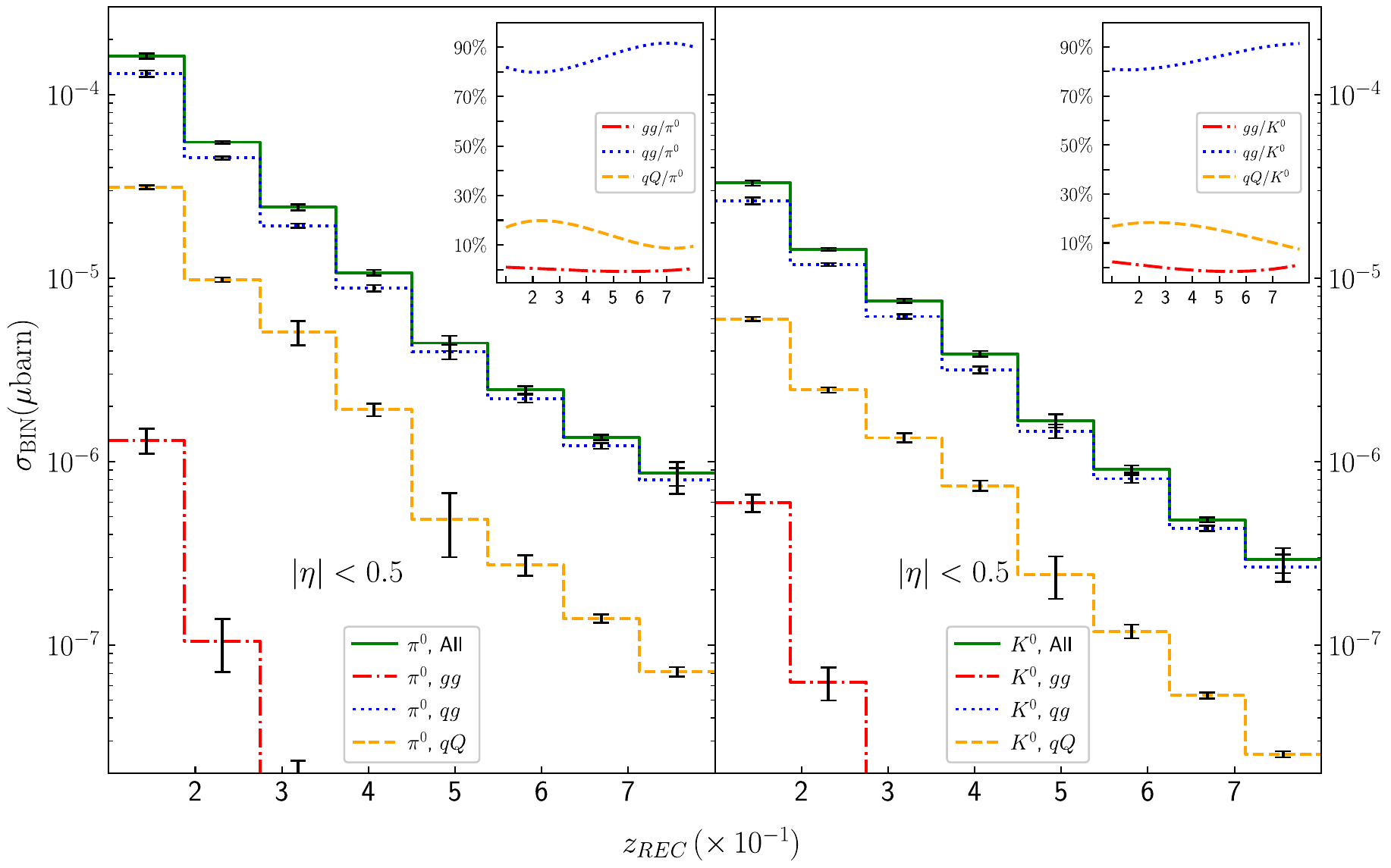}
    \vspace{0.5cm} 
    \includegraphics[width=0.49\textwidth]{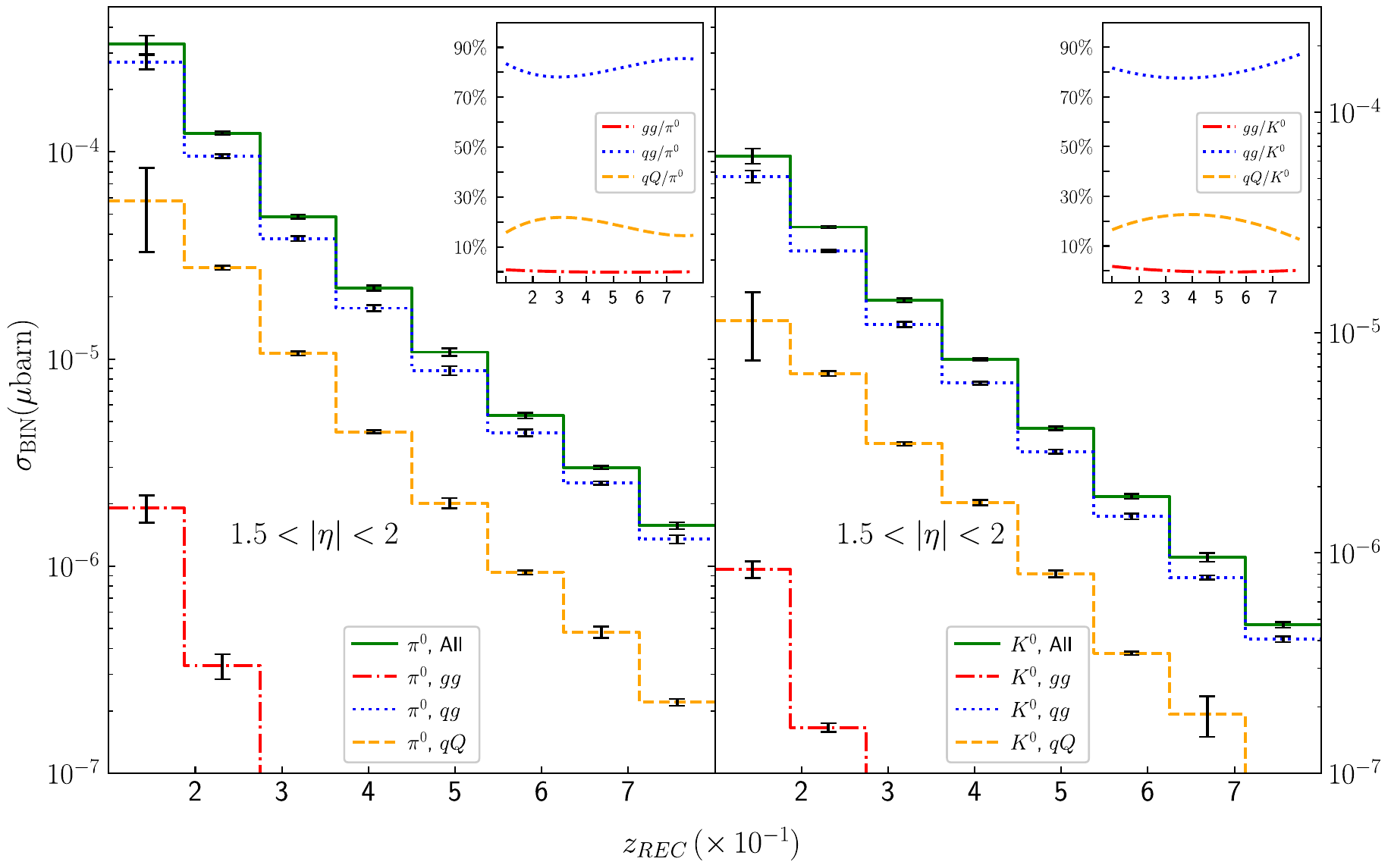}
    \caption{Differential cross-section distribution as a function of $z_\text{REC}$ for neutral hadron production for $\vert\eta\vert < 0.5$ (top) and $1.5<\vert\eta\vert  < 2$ (bottom). We indicate the contributions due to the different partonic channels: $gg$ (dashed red line), $qg$ (point blue line), and $qQ$ (dashed orange line). We use Tevatron kinematics ($\sqrt{S_\text{C.M.}}=1.938\,\text{TeV}$). In the upper sector of each plot, we indicate the relative contribution of each channel.}
    \label{fig:Figura3}
\end{figure}

\begin{figure*}[t!]
    \centering
    \includegraphics[width=0.49\textwidth]{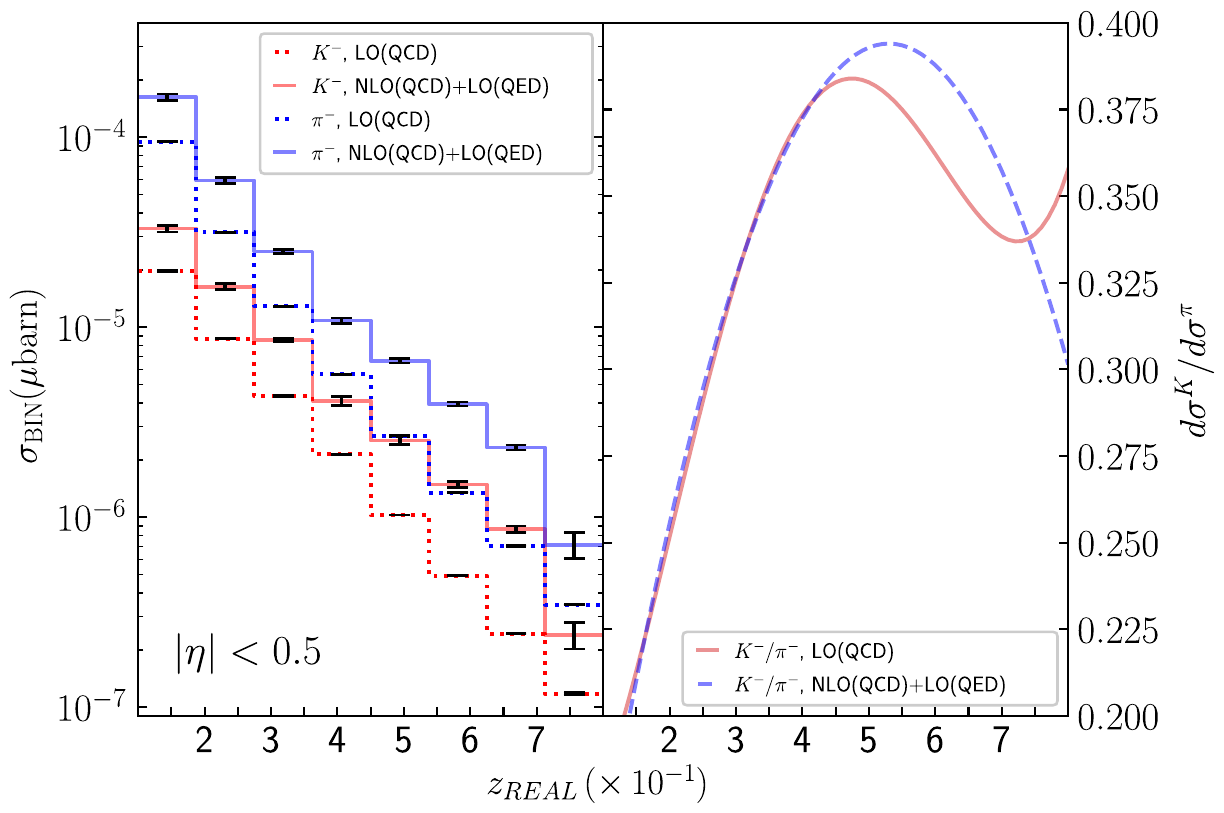}
    \includegraphics[width=0.49\textwidth]{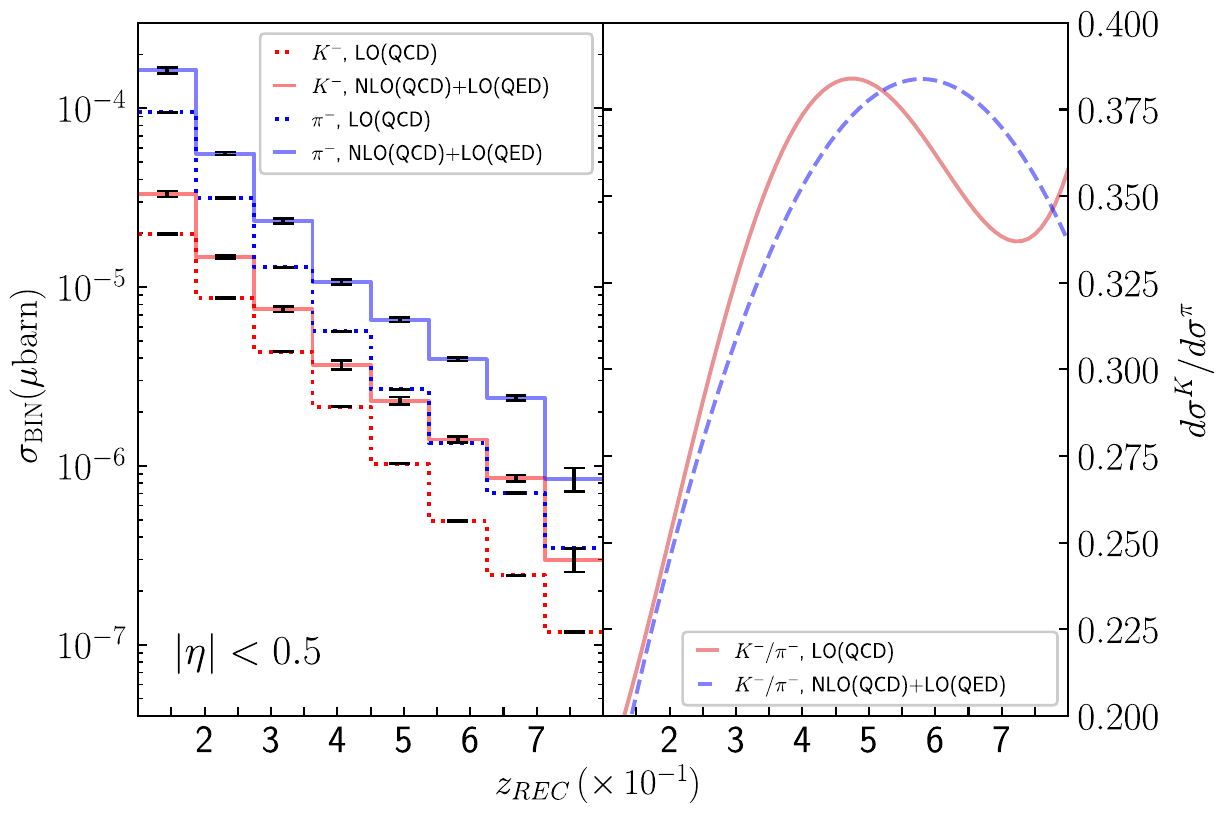}
    \includegraphics[width=0.49\textwidth]{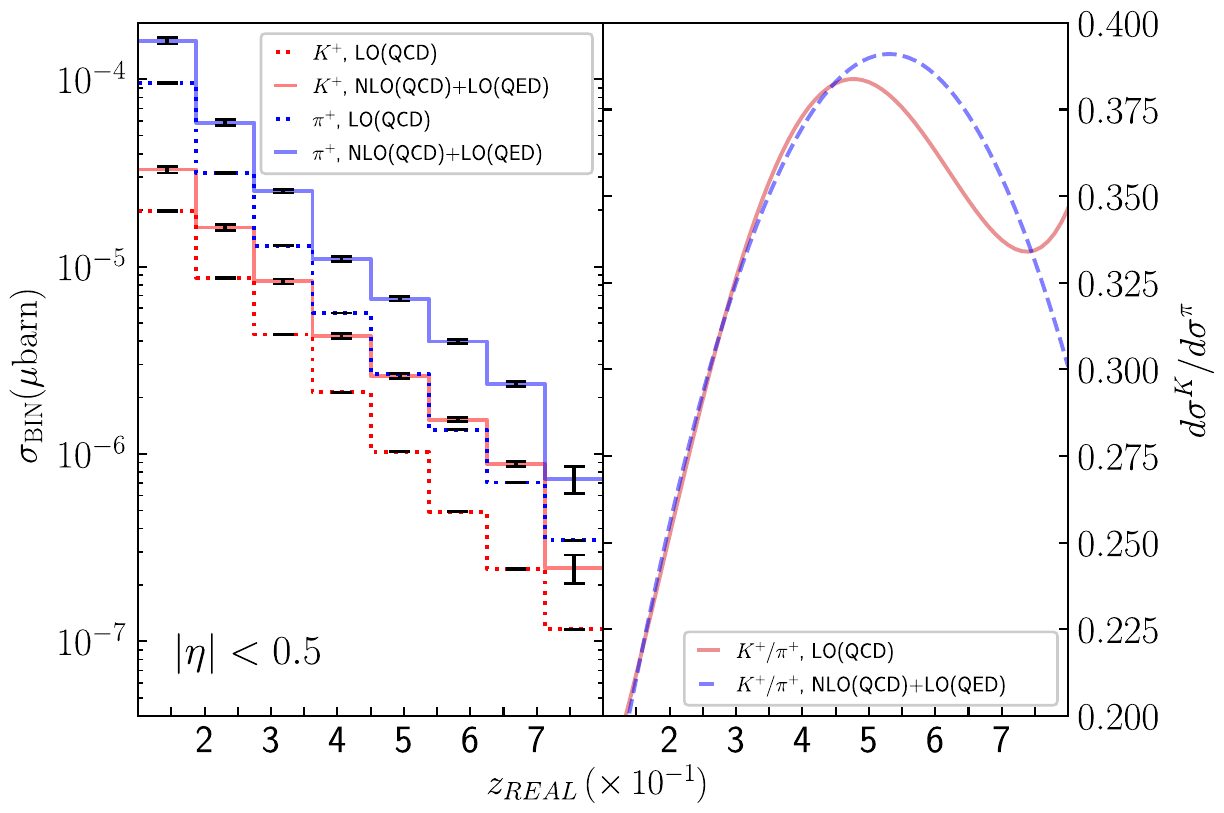}
    \includegraphics[width=0.49\textwidth]{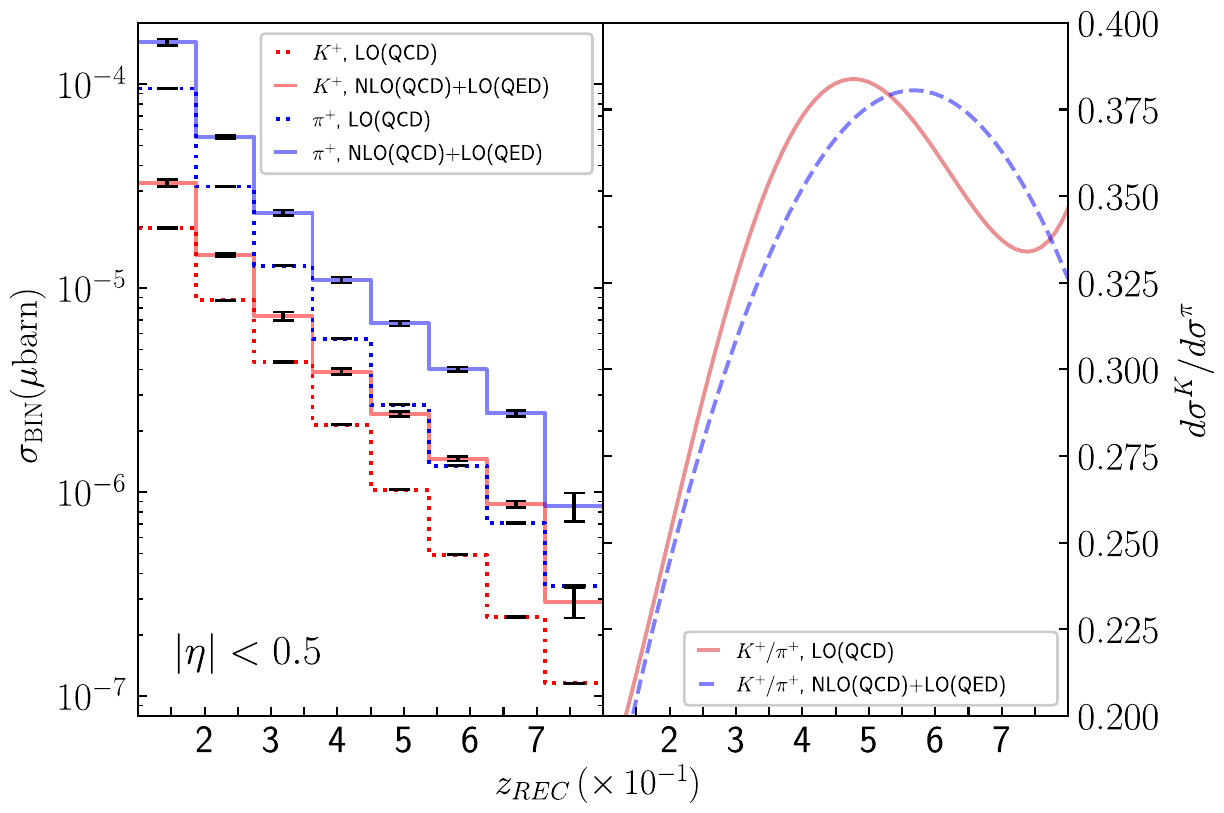}
    \includegraphics[width=0.49\textwidth]{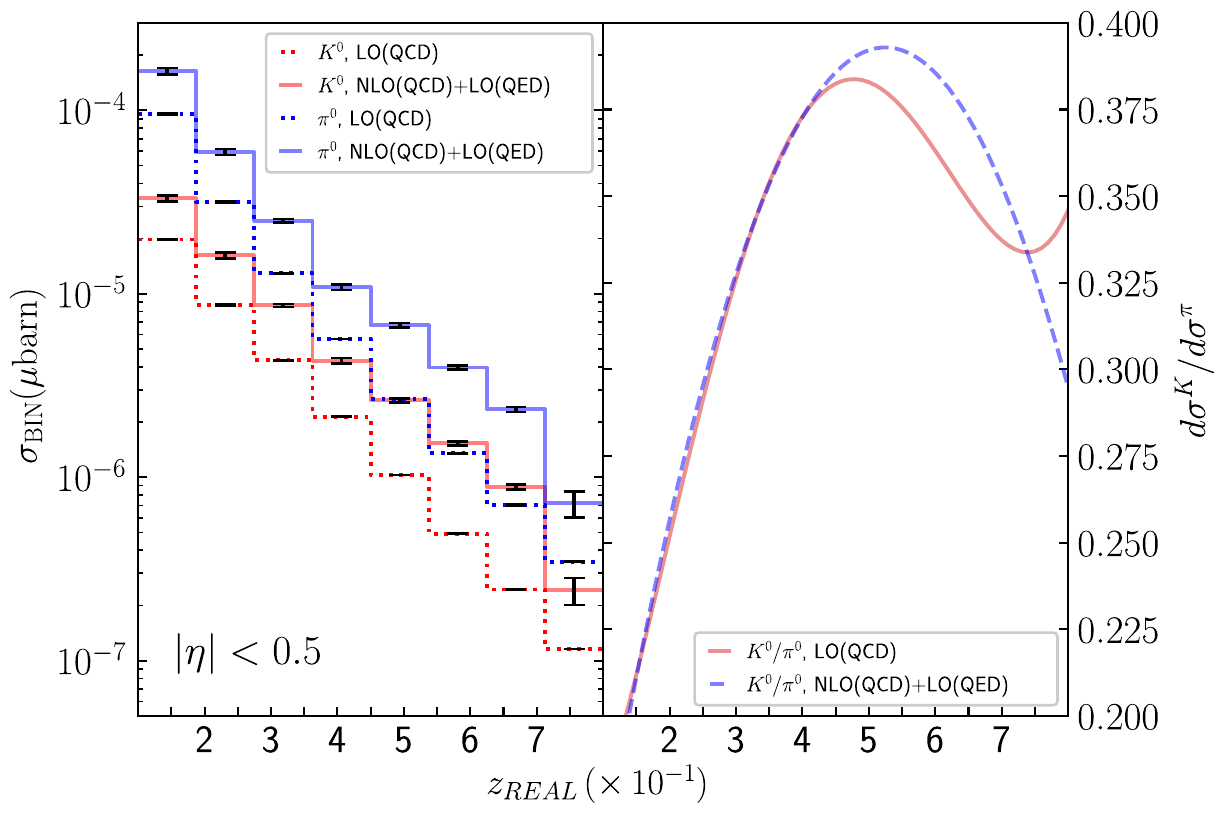}
    \includegraphics[width=0.49\textwidth]{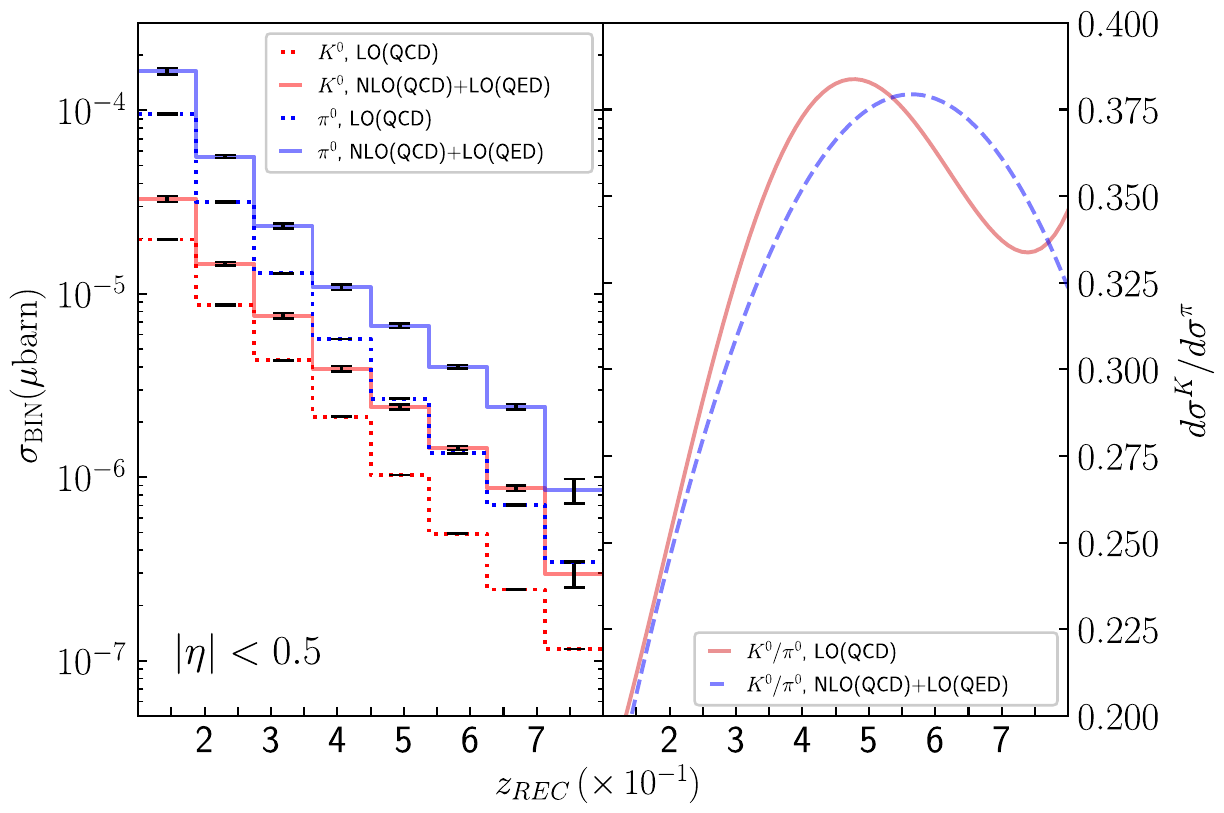}
    \caption{Differential cross-section distribution of $\pi$ and $K$ as a function of $z_{REAL}$ ({left} column) and $z_{REC}$ ({right} column), using Tevatron kinematics ($\sqrt{S_\text{C.M.}} = 1.986 \, {\rm TeV}$). In the left panels, we display the differential cross-sections, whilst in the right ones we present the ratio $d\sigma^K/d\sigma^\pi$. We present the results for negative (first row), positive (second row) and neutral (third row) hadron production for $\vert\eta\vert<0.5$. {This is the {\it Scenario 1} configuration.} }
    \label{fig:Figura4}
\end{figure*}

\begin{figure*}[t!]
    \centering
    \includegraphics[width=0.49\textwidth]{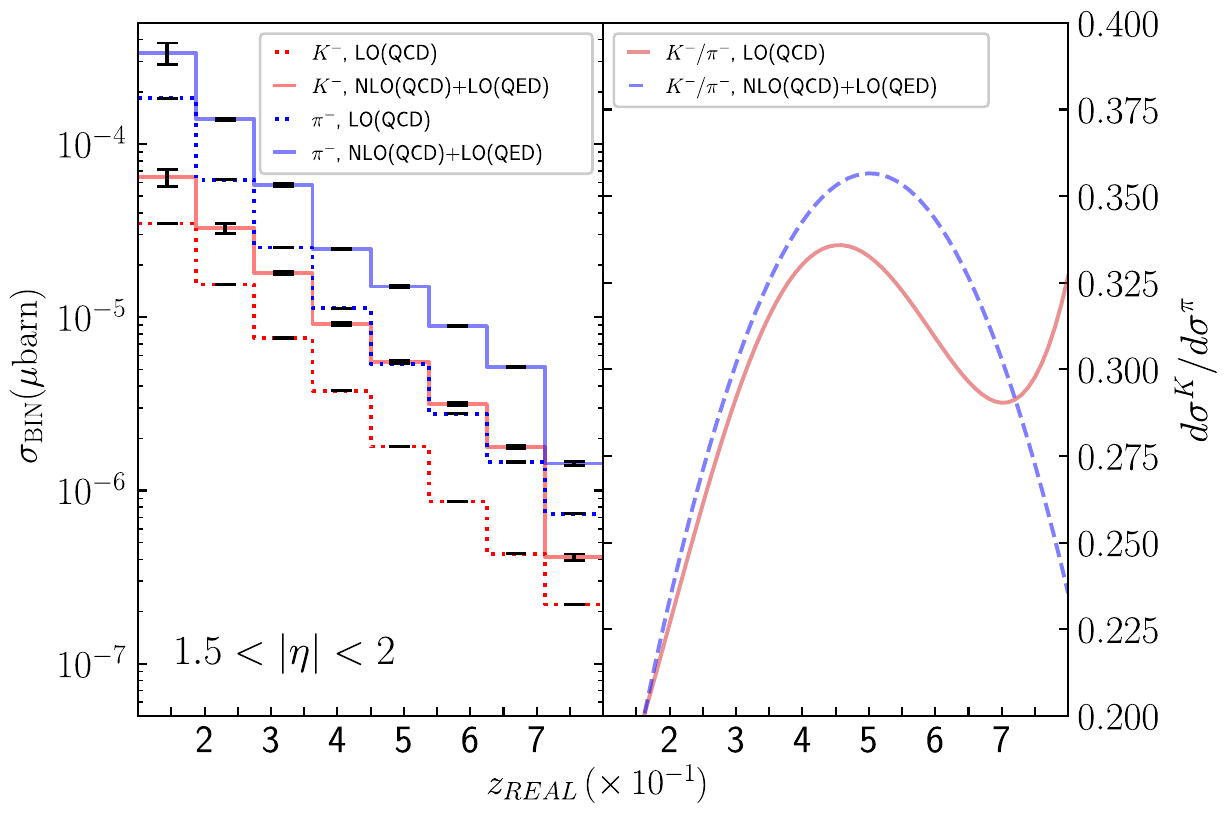}
    \includegraphics[width=0.49\textwidth]{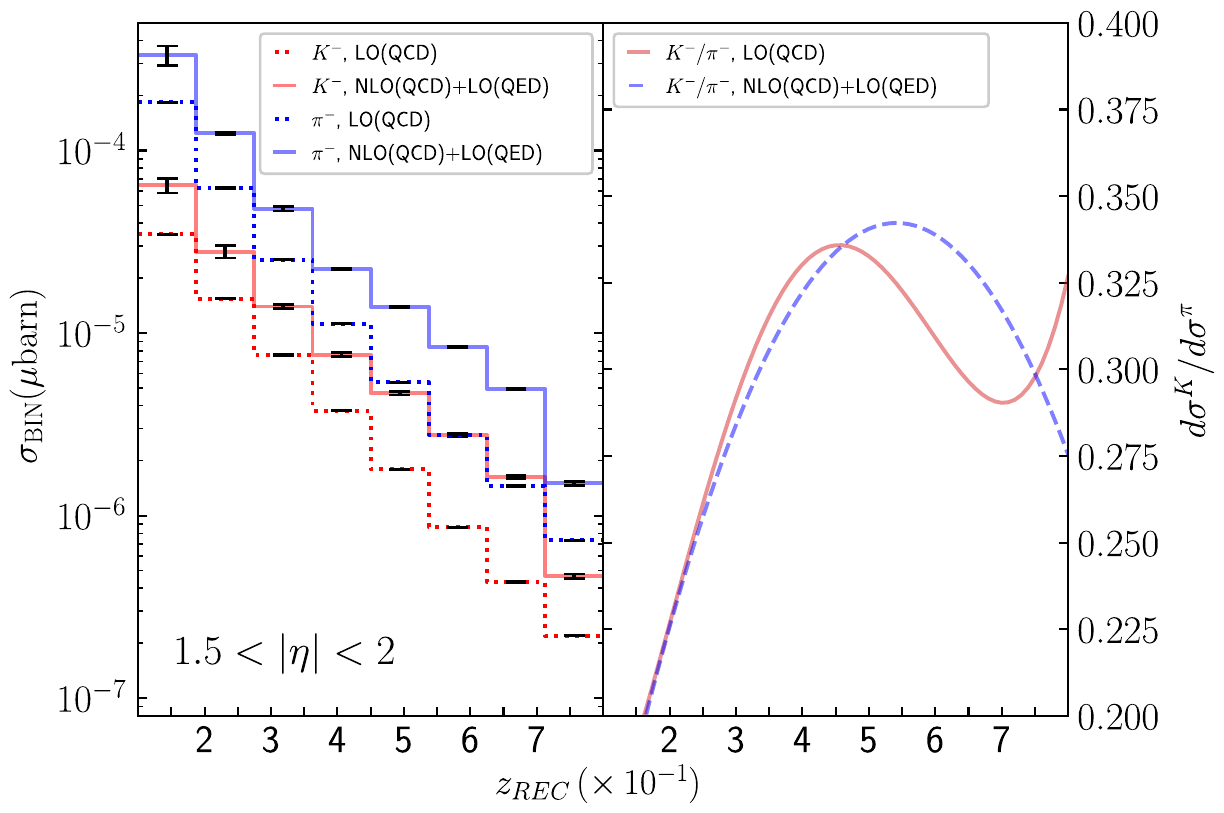}
    \includegraphics[width=0.49\textwidth]{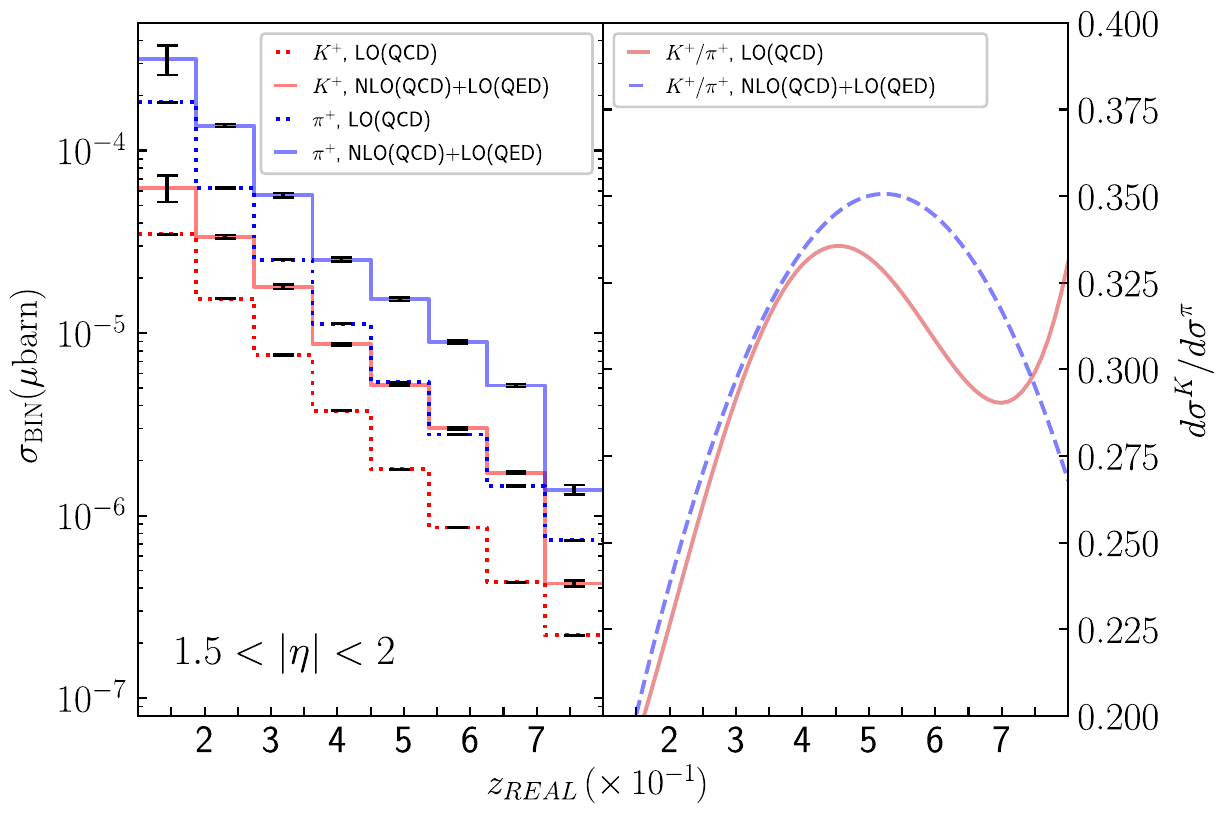}
    \includegraphics[width=0.49\textwidth]{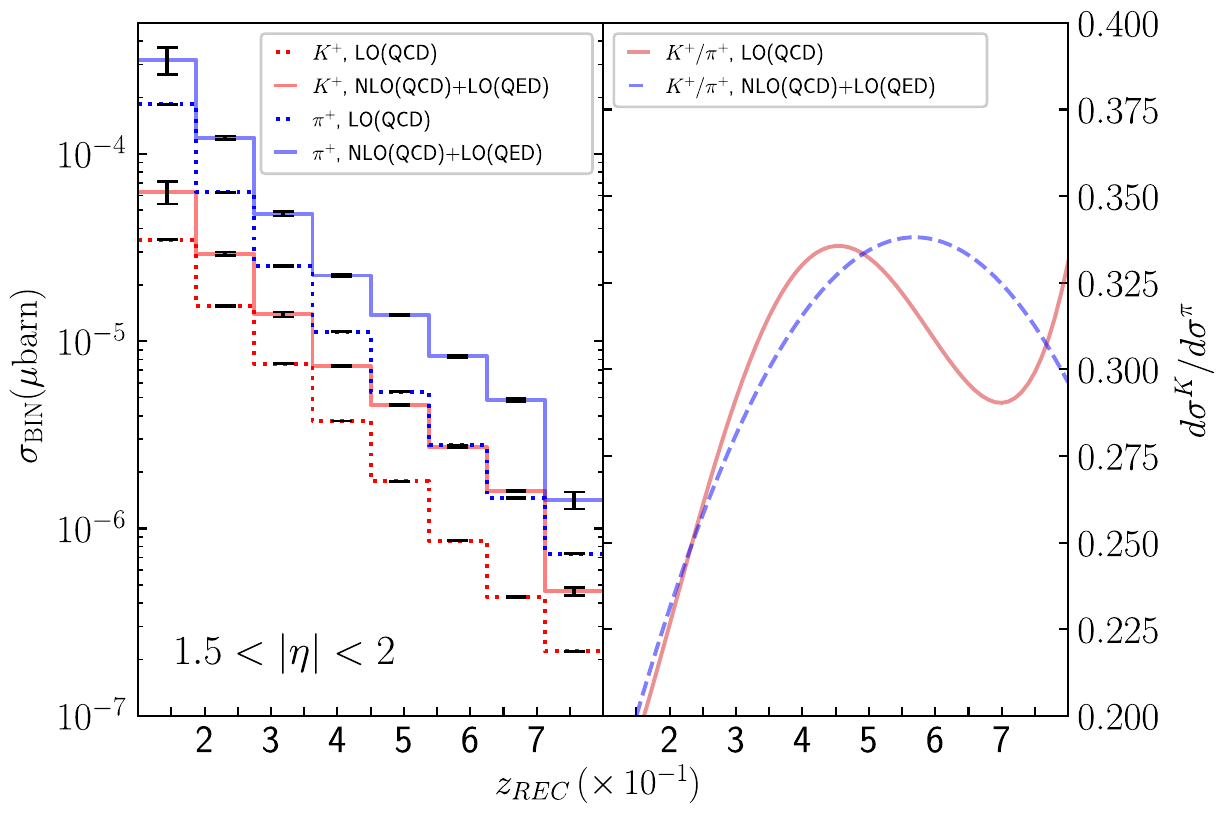}
    \includegraphics[width=0.49\textwidth]{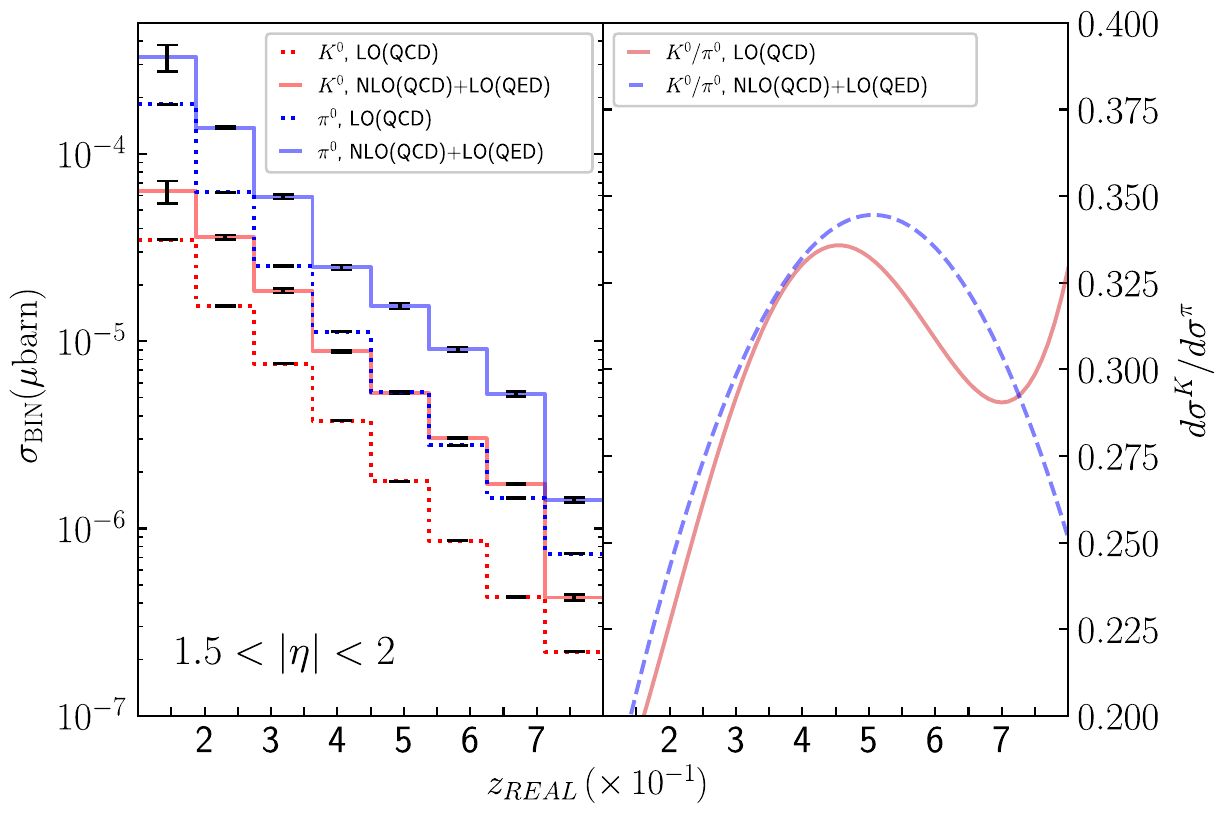}
    \includegraphics[width=0.49\textwidth]{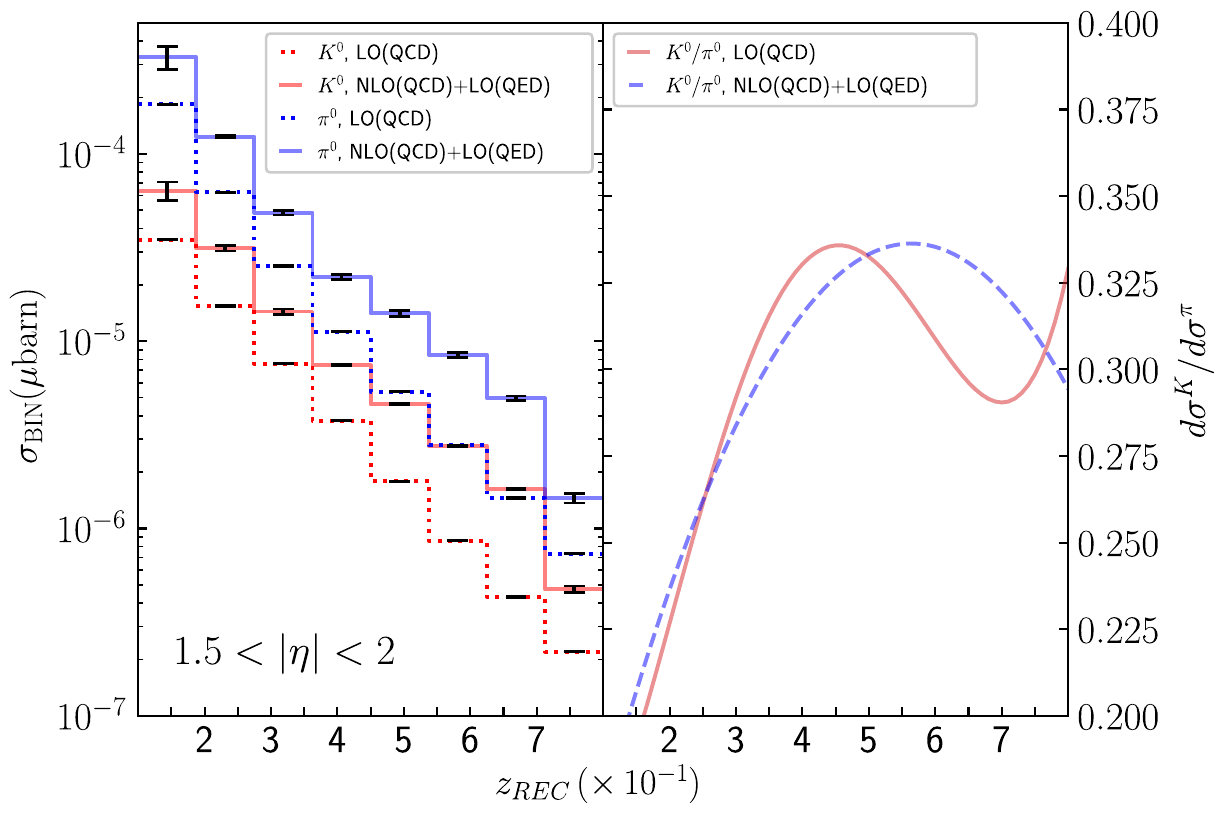}
    \caption{Differential cross-section distribution of $\pi$ and $K$ as a function of $z_{REAL}$ ({left} column) and $z_{REC}$ ({right} column), using Tevatron kinematics ($\sqrt{S_\text{C.M.}} = 1.986 \, {\rm TeV}$). In the left panels, we display the differential cross-sections, whilst in the right ones we present the ratio $d\sigma^K/d\sigma^\pi$. We present the results for negative (first row), positive (second row) and neutral (third row) hadron production for $1.5<\vert\eta\vert<2 $. {This is the {\it Scenario 2} configuration.}}
    \label{fig:Figura5}
\end{figure*}

In Fig.~\ref{fig:Figura3} we show differential cross-section distributions as a function of $z_{\rm REC}$ per channel for {\it Scenario 1} (top) and {\it Scenario 2} (bottom). For the sake of simplicity, we present results of neutral hadron-photon production; nevertheless, we computed the corresponding charged hadron distributions. We distinguish a clear dominance of the $gq$-channel, followed by the $qQ$-channel, where the $gg$-channel is the less dominant one. For completeness, in Fig.~\ref{fig:Figura4} and  Fig.~\ref{fig:Figura5} we present similar plots to Fig.~\ref{fig:Figura1} {(Fig.~\ref{fig:Figura4} refers to {\it Scenario 1} and Fig.~\ref{fig:Figura5} to {\it Scenario 2})}. In general, we find that cross-sections as a function of $z_{\rm REC}$ at NLO(QCD)+LO(QED) are close to the distributions in terms of $z_{\rm REAL}$. Then, we can still rely on the phenomenology with the variable $z_{\rm REC}$.

\begin{figure}[t!]
    \centering
    \includegraphics[width=0.49\textwidth]{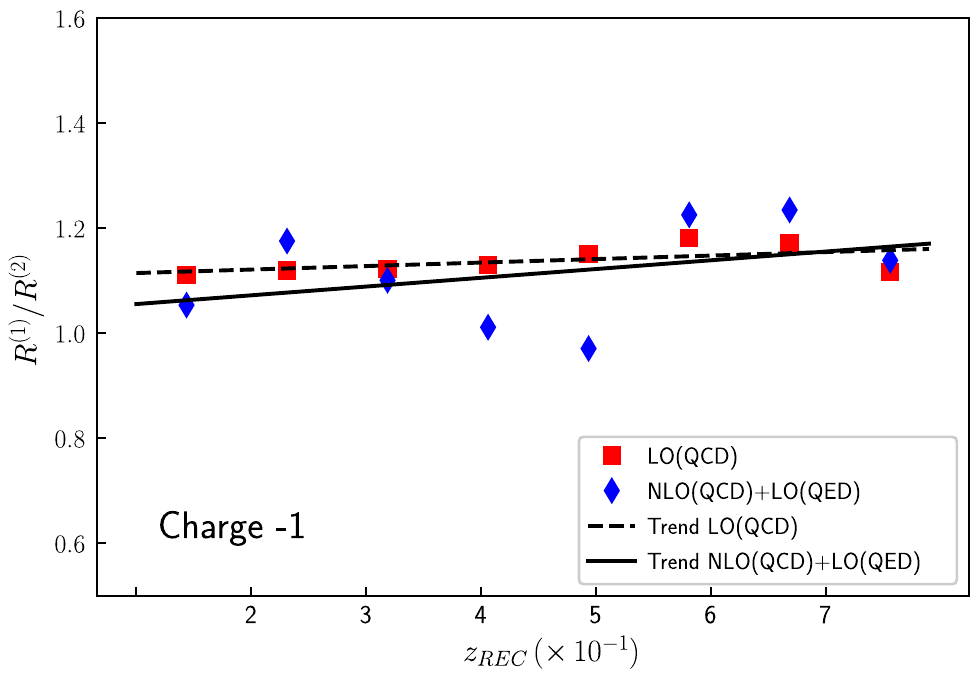}
    \vspace{0.5cm} % espacio vertical entre las figuras
    \includegraphics[width=0.49\textwidth]{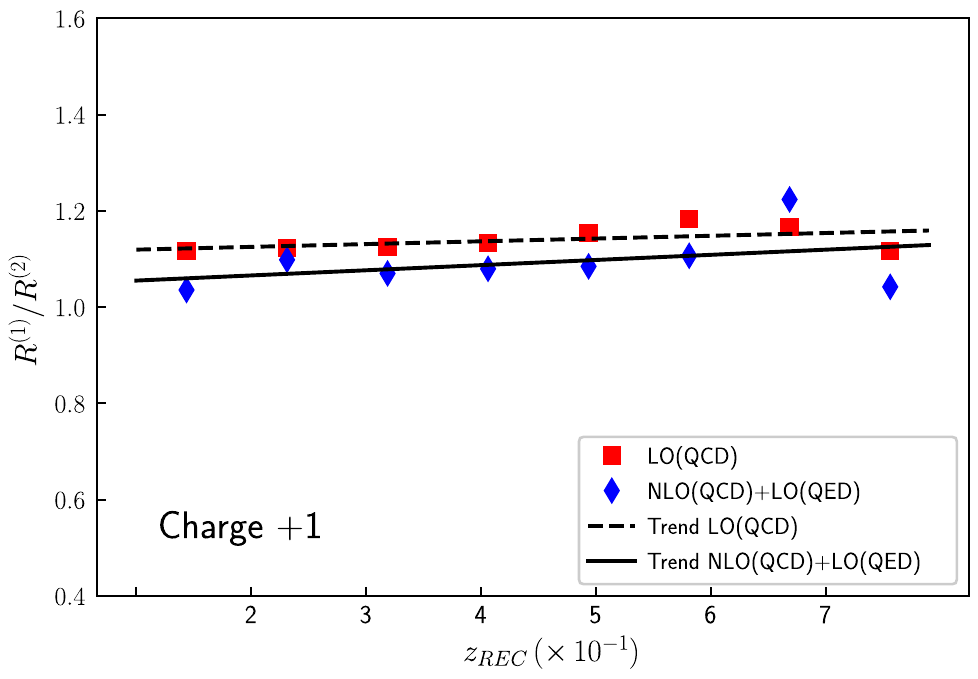}
     \includegraphics[width=0.49\textwidth]{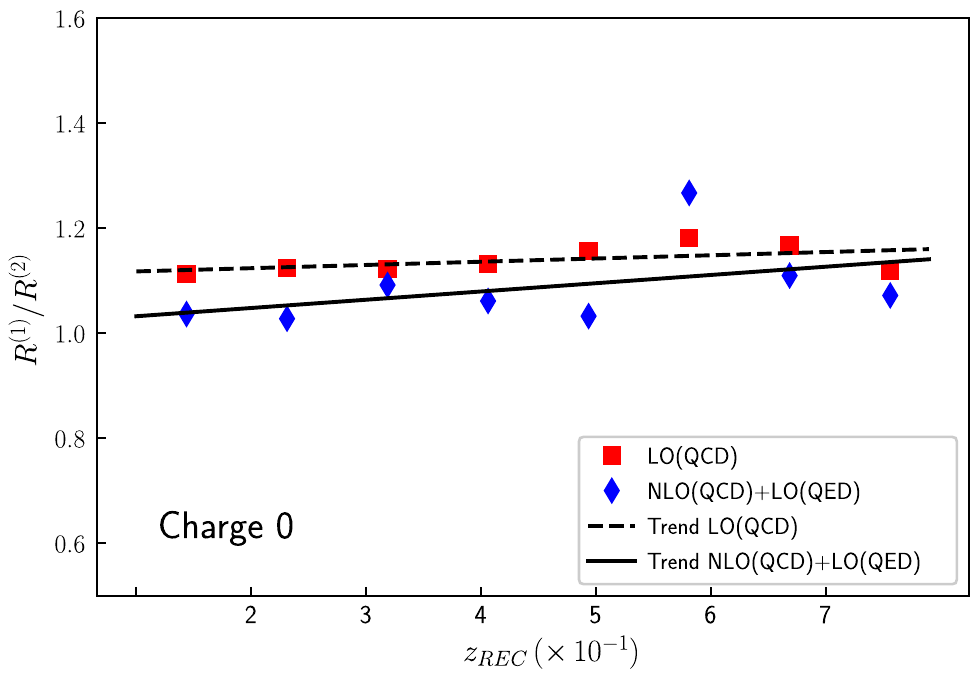}
    
    \caption{Comparison of cross-section ratios $R^{(1)}/R^{(2)}$ as a
function of $z_\text{REC}$. We include the LO QCD (red) and NLO QCD $+$ LO QED (blue) predictions, for negative (top), positive (center), and neutral (bottom) hadron production.}
    \label{fig:Figura6}
\end{figure}

Before concluding this section, we analyze how the differential cross-section distributions and $d\sigma^K/d\sigma^\pi$ ratios behave when the two kinematical cuts are applied.  Results in Fig.~\ref{fig:Figura6} are presented in terms of the ratio $R^{(1)}/R^{(2)}$, which is defined as:
\begin{equation}
R^{(1)}=\left.\frac{d\sigma^K}{d\sigma^\pi}\right\rvert_{|\eta|<0.5},\hspace{0.8cm}\left.R^{(2)}=\frac{d\sigma^K}{d\sigma^\pi}\right\rvert_{1.5<|\eta|<2}
\end{equation}\label{FTH}
this applies to Tevatron energies and to different hadron charge states. The red (blue) markers represent the LO (NLO QCD + LO QED) predictions, while the dotted (solid) black lines show the linear trends of $R^{(1)}/R^{(2)}$ as a function of $z_\text{REC}$. 
{
The results suggest that the inclusion of NLO QCD + LO QED corrections reduces the $R^{(1)}/R^{(2)}$ ratio across all considered cases, for both neutral and charged hadrons. The suppression in the ratio's behavior is on the order of 10\%. This trend deviates from the observations in proton-proton collisions, where symmetric beams tend to induce opposing behaviors between the positive and negative hadron ratios. Furthermore, in our study, we observe a consistently increasing and nearly flat trend, in contrast to the dynamics observed in the proton-proton case.
The similar behavior observed across the three trends is consistent with the results shown in Figs. \ref{fig:Figura4} and \ref{fig:Figura5}, where the NLO QCD + LO QED correction consistently exceeds the LO QED contribution in all cases. This stands in contrast to the phenomena observed in $pp$ collisions
\cite{Ochoa-Oregon:2023ktx}.}
%
%The results suggest that NLO QCD + LO QED corrections reduce the cross section for negatively charged hadrons and increase it for positively charged hadrons, thereby maintaining the ratios at around $\mathcal{O}(1.1)$. This charge-dependent behaviour stems from the $z$-dependence of the FFs, confirming the trends observed in Figures \ref{fig:Figura2} and \ref{fig:Figura3}. For neutral hadrons, the LO and NLO QCD + LO QED contributions tend to cancel each other out, resulting in a similar trend with respect to $z_\text{REC}$.
%%%%%%%%%%%%%%%%%%%%%%%%%%%%%%%%%%%%%%%%%%
%%%%%%%%%%%%%%%%%%%%%%%%%%%%%%%%%%%%%%%%%%

%%%%%%%%%%%%
%%%%%%%%%%%%
\section{Imposing constraints on FFs through cross-section ratios}
\label{sec:FFconstraints}
At this point, it is useful to recall the expression for the hadronic cross section as defined by the factorisation theorem. For the specific case analysed in this study, the result is as follows:
\begin{align}\nonumber
    \label{FTH}
    d\sigma^h&=\sum_{a_1,a_2,a_3}dx_1\, dx_2\, dz\, f_{a_1}^ {H_1}(x_1)\, f_{a_2}^{H_2}(x_2)\, d_{a_3}^{h}(z) \\
       &\times d\hat{\sigma}_{a_1a_2\rightarrow a_3\gamma}(x_1P_1,x_2P_2,P_h/z,P_\gamma).
\end{align}
%%%%%
Here, $\{P_1,P_2\}$ represent the momenta of the incident hadrons $H_1$ and $H_2$, respectively. $P_\gamma$ corresponds to the photon momenta and $P_h$ to the momenta of the produced hadron. The center-of-mass energy is defined as $E_\text{C.M.} = \sqrt{2P_1\cdot P_2}$. All parton contributions, including phase space integrals, higher-order corrections and associated measure functions, are grouped into $d\hat{\sigma}_{a_1 a_2\rightarrow a_3\gamma}$. For simplicity, the explicit dependence on the factorization scale in the PDFs and FFs is omitted. Next, we claim that when kinematic cuts are applied, the selection of events is such that differential cross-section over $z$, can be rewritten as
\begin{align}\nonumber
\label{dxg}
    \frac{d\sigma^{h_i}}{dz}=&\sum_{a_3}d_{a_3}^{h_i}\\\nonumber
    \times&\left[\sum_{a_1a_2}\int dx_1dx_2 f_{a_1}^{H_1}(x_1)f_{a_2}^{H_2}(x_2)d\hat{\sigma}_{a_1a_2\rightarrow a_3\gamma}\right]\\
    =&\sum_{a_3}d_{a_3}^{h_i}\times g_{a_3}(z);
\end{align}
and $h_i$ represents the identified hadron. In this context, we introduce a new function, $g$, which is independent of the identity of the hadron produced in the final state.
{We must emphasize that this expression remains valid for any calculation within the framework of pQCD. The objective of reformulating it in this manner is to highlight that, should it be possible to isolate channel-by-channel contributions experimentally, the determination of fragmentation function ratios could be estimated with greater ease. To demonstrate that a partial selection of leading terms is achievable, the previous analyses allows us to conclude that:}
\begin{itemize}
\item the variable $z=z_\text{REAL}$ exhibits a behavior closely resembling that of $z_\text{REC}$, and \
\item a single partonic channel contribution provides the dominant effect.
\end{itemize}

Under the first assumption, inspired by Fig.~\ref{fig:Figura1}, Fig.~\ref{fig:Figura4} and Fig.~\ref{fig:Figura5}, $z \rightarrow z_\text{REC}$ can be replaced without significantly affecting the factorization, provided $z_\text{REC}$ and $z_\text{REAL}$ are strongly correlated, especially in certain kinematic regions \cite{Renteria-Estrada:2022scipost}. Regarding the second assumption, motivated by Fig.~\ref{fig:Figura2} and Fig.~\ref{fig:Figura3} and discussed in Sec.~\ref{ssec:DistributionsZparton}, applying appropriate kinematic cuts enables specific parton channels to be enhanced or suppressed. In the analyzed case, it was observed that the $qg$ channel could be amplified to exceed the other contributions by more than an order of magnitude. Additionally, it can be assumed that the $u$ channel dominates. This is justified not only because protons contain more up quarks than down quarks, but also because at amplitude level,
\begin{equation}
\label{eq:amplitudes}
    |\mathcal{M}_{ug\rightarrow u\gamma}|^2=4|\mathcal{M}_{dg\rightarrow d\gamma}|^2,
\end{equation}
which is a direct consequence of photon production. Taking this into account and considering that we can focus primarily on the kinematics where the $qg$-channel is dominant, we conclude that
\begin{equation}
\label{Rkpi}
R^{K/\pi}(d\sigma)=\frac{d\sigma^K/dz_\text{REC}}{d\sigma^\pi/dz_\text{REC}}\approx \frac{d^K_u(z_\text{REC})}{d^\pi_u(z_\text{REC})}=R^{K/\pi}(d_u) \, ,
\end{equation}
is a sufficiently accurate approximation. 
{The validity of Eq.~(\ref{Rkpi}) will be evaluated below. For simplicity, higher-order corrections are limited to NLO QCD as the available FFs do not include terms beyond this level. Furthermore, as the effect of these corrections on the $p+\bar{p}\to\gamma+h$ cross section are expected to be less than $1\%$ \cite{Catani:2002ny}, it is enough to verify Eq.~(\ref{Rkpi}) with NLO QCD precision. Specifically, beyond the NLO order in QCD, the expected corrections for this process are on the order of 1\% or less \cite{Catani:2002ny}; therefore, our current framework remains theoretically robust for the targeted observables.
}
%The validity of Eq.~(\ref{Rkpi}) will be evaluated below. For simplicity, higher-order corrections are limited to NLO QCD as the available FFs do not include terms beyond this level. Furthermore, as the effect of these corrections on the $p+\bar{p}\to\gamma+h$ cross section is less than $1\%$, it is enough to verify Eq.~(\ref{Rkpi}) with NLO QCD precision.

Finally, we emphasize that the assumptions adopted here are independent of the specific nature of $h_1$ and $h_2$. In other words, the comparison holds for any pair of hadrons, regardless of their electromagnetic charge, isospin, or quark content. Deviations from Eq.~(\ref{Rkpi}) are expected when $h_1$ and $h_2$ differ significantly in composition, which motivates our choice of comparing hadrons with the same electric charge. In addition, the presence of different valence quarks alters the relative weight of other initial-state partonic contributions, a clear enhancement of the $u$-initiated FFs is still expected. This explains the similarity observed between $\gamma+\pi^\pm$ and $\gamma+K^\pm$ final states, as well as the slightly larger discrepancies between $\gamma+\pi^0$ and $\gamma+K^0$.

%%%%%%%%%%%%%%%%%%%%%%%%%%%%%%%%%%%%%%%%%%
%%%%%%%%%%%%%%%%%%%%
%%%%%%%%%%%%%%%%%%%
\subsection{Cross-section vs. FFs ratios}
\label{ssec:Comparison}
 %The first step in testing our approximations is the calculation of the FFs ratio [rhs of Eq.~\ref{Rkpi}]. Since FFs are energy-dependent, it is essential to select a physically meaningful reference value. Moreover, this reference energy must be connected, in some way, to the physical cross sections. As discussed in Sec.~\ref{sec:Analysis}, the default scale choice, $\mu$, is defined as the average of the transverse momenta of the produced hadron and the photon. According to Eq.~\ref{mu}, this quantity depends on each individual event. While this choice is fully appropriate for cross-section computations, it is not suitable for FFs determination. In fact, FFs are first defined at a fixed scale and subsequently evolved through the DGLAP equations [13,66,77]. To address this issue, we define the average energy scale, $\bar{Q}$, as}
The first step towards the verification of the validity of our assumption in Eq.~(\ref{Rkpi}), is the computation of FFs ratio. Since FFs depend on energy, it is necessary to choose a physically relevant reference value related to cross sections. As explained in Sec.~\ref{sec:Analysis}, the default scale $\mu$ is defined as the average of the transverse momenta of the hadron and photon produced. {Since $\mu=p_T^\gamma$}, this quantity varies in each event, which is suitable for cross section calculations but not for determining FFs, since these are initially defined at a fixed scale and then evolved. % using the DGLAP equations \cite{Altarelli:1977zs,Dokshitzer:1977sg,Gribov:1972ri}. 
To resolve this issue an average energy scale per bin, $\bar{Q}$, is introduced. This quantity is defined as
\begin{eqnarray}
\label{Qprom}
\bar{Q}=\frac{1}{\sigma_{\rm T}}\sum_i\mu_i(p_T^\gamma,p_T^h)(\sigma_\text{BIN})_i
\end{eqnarray}
where $(\sigma_{\rm{BIN}})_i$ is the cross-section per $i$-bin and $\sigma_{\rm T}$ is the total cross-section of the process. All information needed to define $\bar{Q}$ was obtained from the histograms generated by the MC code (run with the $\mu$ scale set to its default value). For all configurations analysed in previous sections, the scale parameter $\bar{Q}$ is found to be $25.38\,\text{GeV}$. This establishes an energy scale for the MC simulation, ensuring physical consistency and that the FFs are well defined at the $\bar{Q}$ scale.
 %%%
 \begin{figure}[t!]
    \centering
    \includegraphics[width=0.49\textwidth]{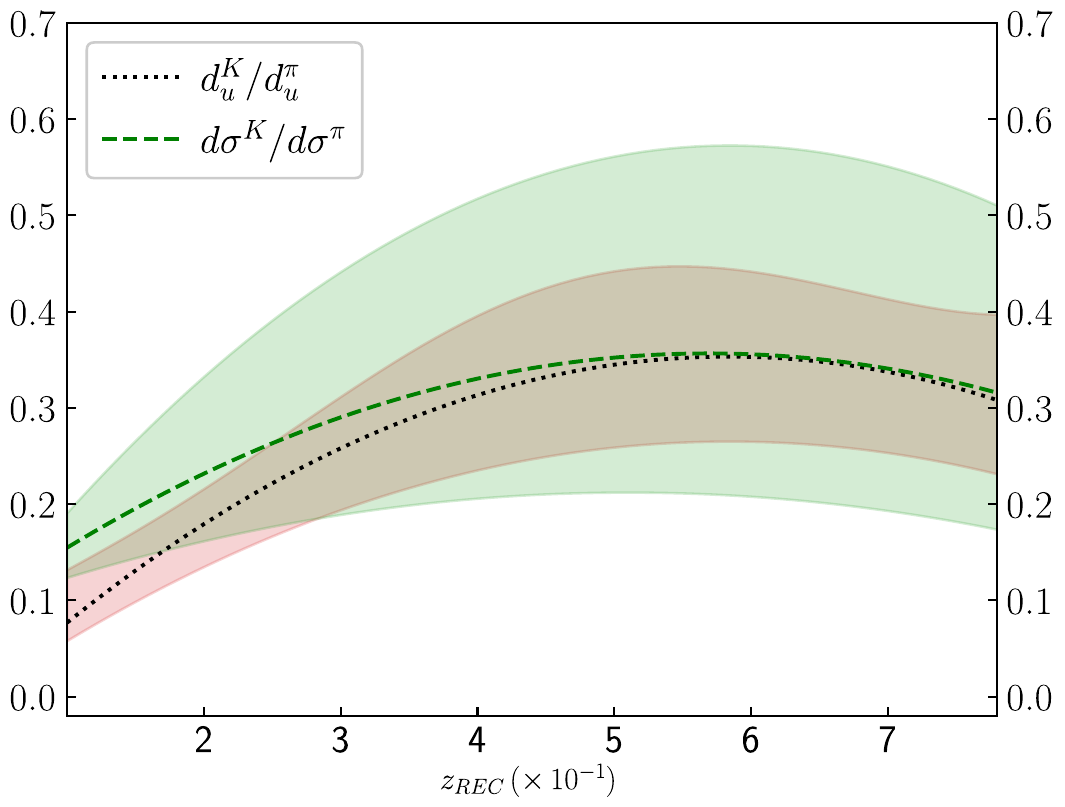} 
    \includegraphics[width=0.49\textwidth]{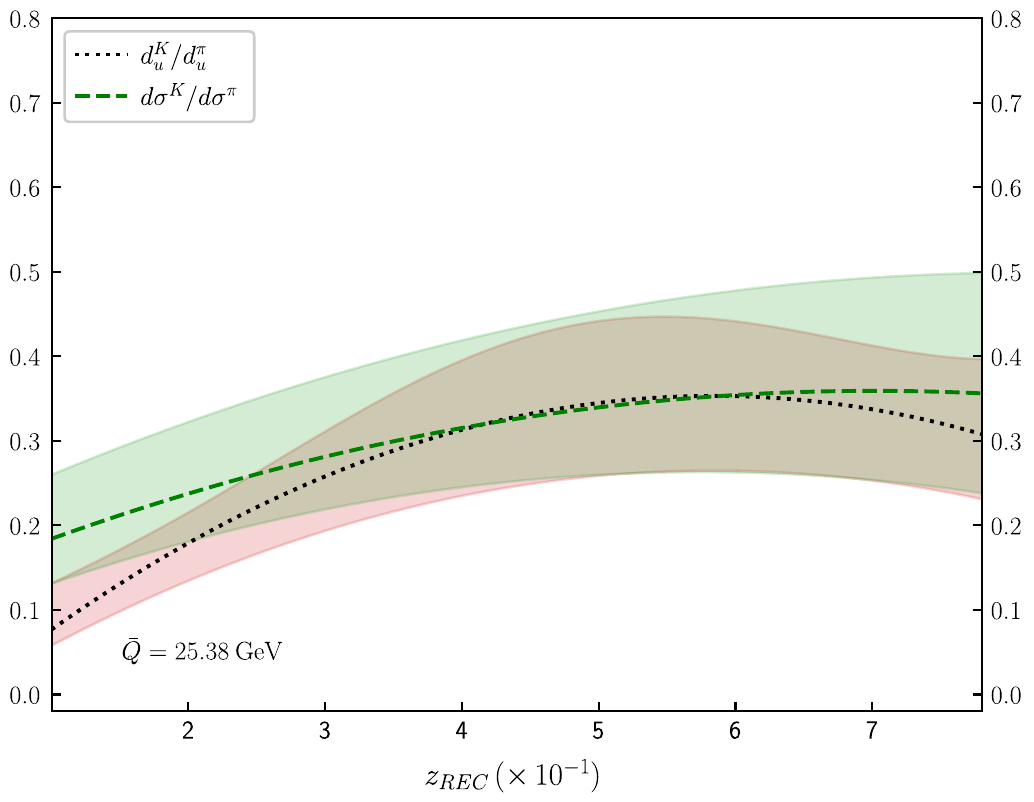} 
\caption{Comparison of $u$-started FF ({dotted} black line) and cross-section ratios (green dashed line) as a function of $z_\text{REC}$, for positive kaon w.r.t. pion production up to NLO QCD accuracy. For the cross-section calculation, we consider two scenarios: (1) with the default energy scale (upper plot), and (2) fixing the reference scale to $\bar{Q}=25.38\, \text{GeV}$ (lower plot). Error bands were computed by varying energy scales by a factor of one-half and two.}
    \label{fig:Figura7}
\end{figure}

In Fig.~\ref{fig:Figura7} a comparison of FF ratios and cross sections as a function of $z_\text{REC}$ for positive hadron production is presented. As discussed in the previous section, the analysis is restricted to fragmentation processes initiated by up quarks. Consequently, $R^{K/\pi}(d_u)$, evaluated at the reference scale $\bar{Q}=25.38,\text{GeV}$ and including NLO QCD corrections, is shown as the dotted black line. As can be seen, the correlation between $d\sigma^K/d\sigma^{\pi}$ and $d_u^K/d_u^{\pi}$ is stronger when $\bar{Q}$ is used. This claim is because there is greater overlap in the error bands when fixing $\bar{Q}$ rather than using the default energy scale.

%%%%%%%%%%%%%%%%%%%%%%%%%%%%%%%%%%%%%%%%%%
\subsection{Improved analysis with physically-motivated cuts}
\label{ssec:FFimproved}
\begin{figure}[t!]
    \centering
    \includegraphics[width=0.49\textwidth]{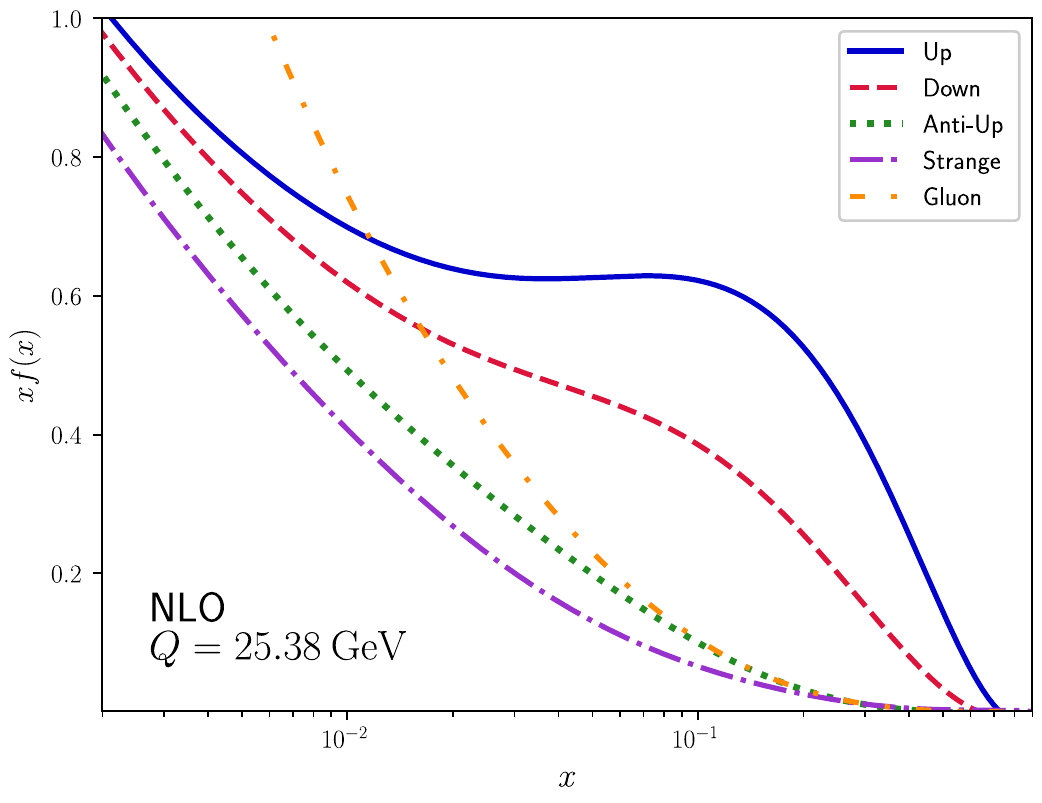} 
    \caption{ The parton density functions (PDFs) were extracted from the \texttt{NNPDF31\_nlo\_as\_0118\_luxqed} dataset. These were evaluated at a scale energy of $\bar{Q}=25.38\,\text{GeV}$. Only the up, down, anti-up and strange flavors were considered, since they contribute mostly to the total cross section of this event. Finally, the gluon PDFs were divided by a factor of 10 to align with the other distributions. }
    \label{fig:Figura8}
\end{figure}

As discussed previously, the two scenarios revealed that $R^{K/\pi}(d_u)$ and $R^{K/\pi}(d\sigma)$ have remarkably similar shapes, with considerable overlap in their uncertainty bands. However, slight discrepancies in the central values are evident due to contributions from different parton channels (PDFs) during the collision and in the hadronization process (FFs). To refine the analysis, we can reinforce the dominance of processes initiated by $u$ quarks by examining the behavior of the PDFs. To this end, we can consider the distributions from the \texttt{NNPDF31\_nlo\_as\_0118\_luxqed} set, shown in {Fig.~\ref{fig:Figura8}}, which demonstrate that the $u$ and $d$ PDFs diverge significantly within the $x$ range of $(0.05,0.3)$. Within this range, the $u$-quark PDF is approximately $50\%$ larger than the $d$-quark PDF and almost three times larger than the $s$-quark PDF. The other quark flavors are even more suppressed compared to the $u$-quark PDF, while $f_g$ is about an order of magnitude larger than any quark PDF. This implies that the luminosity of the $u-g$ channel is at least $50\%$ higher than that of the $d-g$ channel within the specified range, and over an order of magnitude higher than the other contributions. Based on this observation, a cut-off has been implemented to restrict the range of $x$, since $x$ cannot be measured directly. The discussion in Refs. \cite{deFlorian:2010vy,Renteria-Estrada:2022scipost} is followed, and an approximation in terms of experimentally accessible quantities is employed. {Motivated by the LO kinematics, we define} the reconstructed momentum fractions as
\begin{align}
(x_1)_\text{REC}&=p_T^\gamma\frac{\exp(\eta^\pi)+\exp(\eta^\gamma)}{\sqrt{S_{C.M.}}}, \\
(x_2)_\text{REC}&=p_T^\gamma\frac{\exp(-\eta^\pi)+\exp(-\eta^\gamma)}{\sqrt{S_{C.M.}}},
\end{align}
which are strongly correlated with the real MC partonic momentum fractions over a wide kinematical range \cite{Renteria-Estrada:2022scipost}.

\begin{figure}[t!]
    \centering
    \includegraphics[width=0.49\textwidth]{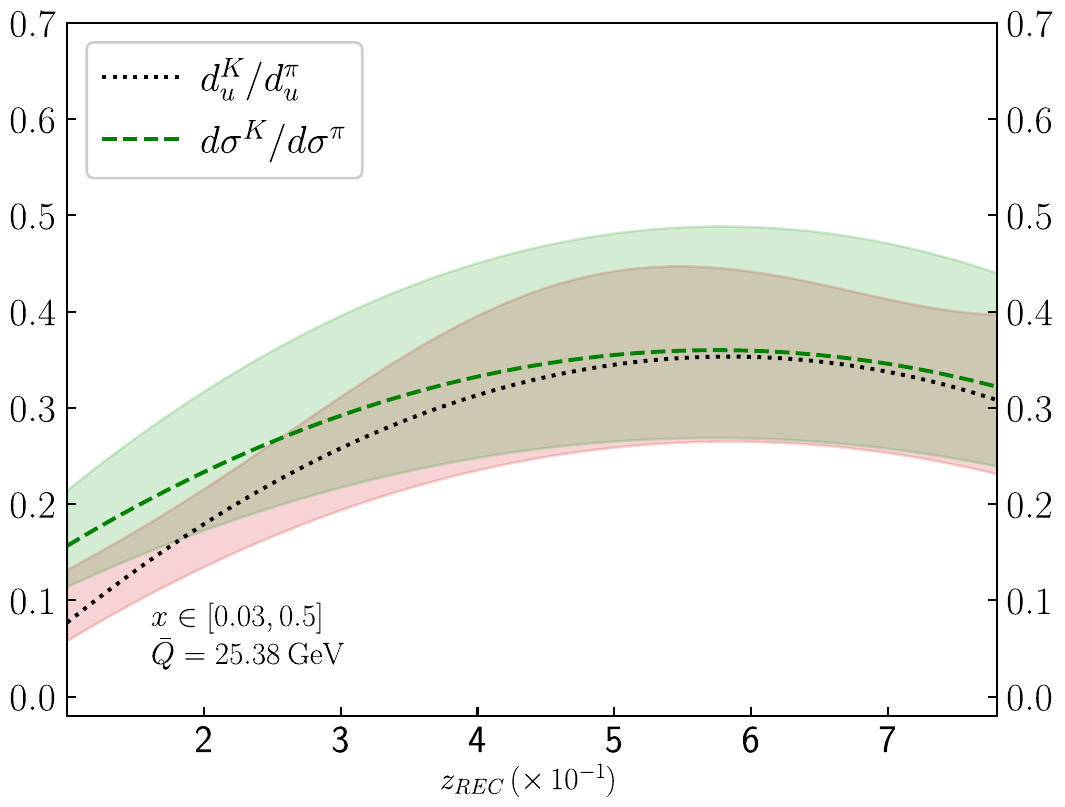} 
    \caption{The figure shows a comparison of the $u$-initiated FFs (black dashed line) and the cross-section ratios (green dashed line) as a function of $z_\text{REC}$ for the production of positive kaons and pions. These were computed up to NLO accuracy in QCD. For the cross-section evaluation, the reference scale is set to 25.38 GeV, and the momentum fraction of the colliding partons is limited to the range (0.03, 0.5), thereby enhancing the $u$-channel contribution. Error bands were computed by varying energy scales by a factor of one-half and two.}
    \label{fig:Figura9}
\end{figure}

\begin{figure}[t!]
    \centering
    \includegraphics[width=0.49\textwidth]{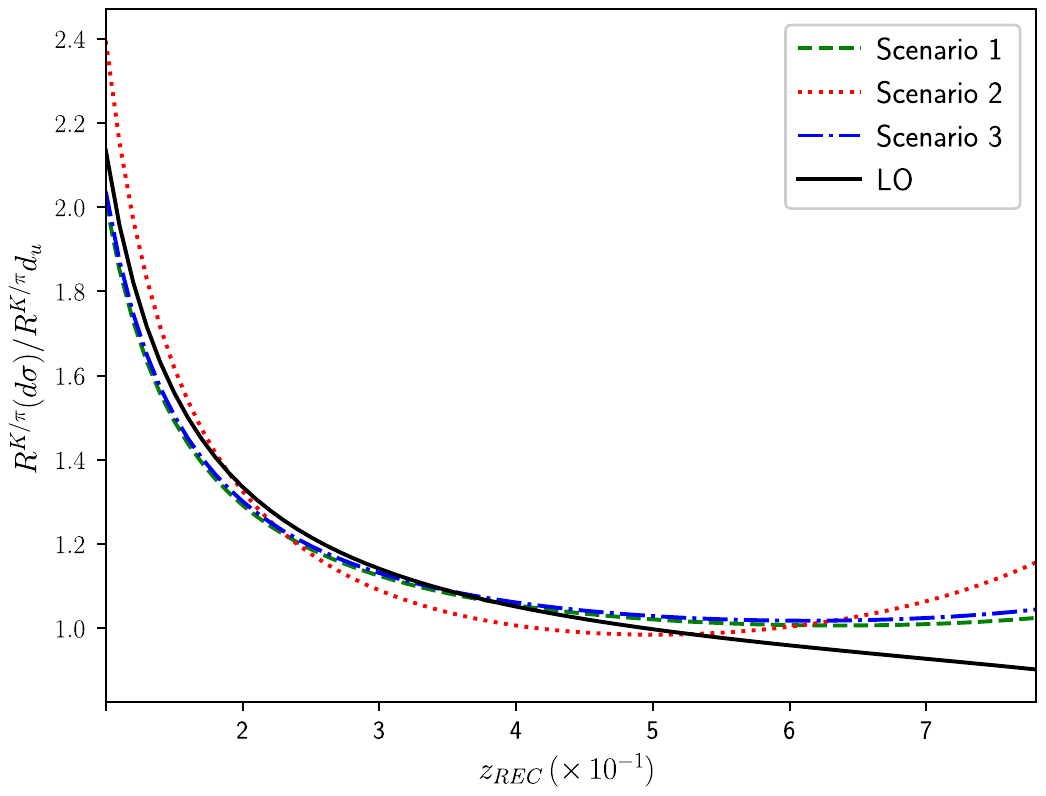} 
    \caption{Ratio $R^{K/\pi}(d\sigma)/R^{K/\pi}(d_u)$, as a function of $z_\text{REC}$ for {\it Scenario~1} (green), {\it Scenario~2} (red) and {\it Scenario~3} (blue). For comparison, we include the LO (black) ratio for {\it Scenario~3}, to estimate the impact of the high order corrections. }
    \label{fig:Figura10}
\end{figure}

We then define {\it Scenario 3} by imposing cuts,
\begin{equation}
    0.03\le\{(x_1)_\text{REC},(x_2)_\text{REC}\}\le 0.5,
\end{equation}
in addition to the ones applied in {\it Scenario 2}. {It is important to mention that this cut does not impose an additional sharp constraint on the distribution variables; it serves solely to select events, while the full range of Eq.~(\ref{eq:cuts}) remains unaffected.} The results, along with the corresponding error bands calculated using the procedure described above, are presented in Fig. \ref{fig:Figura9}. It can be seen immediately that $R^{K/\pi}(d_u)$ and $R^{K/\pi}(d\sigma)$ are closer than in previous scenarios. In the range $z\in(0.5,0.8)$, we can see that $R^{K/\pi}(d_u)$ and $R^{K/\pi}(d\sigma)$ are closer to each other than in the previous scenarios. Both ratios show similar behaviour, but there is also a deviation in the lower regions of $z$. In any case, the error bands overlap almost throughout the entire range of $z$ considered. This suggests that stronger restrictions can be imposed on $R^{K/\pi}(d_u)$ using the cross section ratio, paving the way for the accurate determination of heavy meson FFs from experimental data.
%Por ultimo, comparamos los tres escenarios propuestos entre ellos, y con la estimación a LO bajo el escenario tres. Como ya explicamos, el canal $qg$ es el canal dominante para la producción de fotón-hadrón en colisionadores, donde $qQ$ es casi un orden de magnitud mas pequeño. A LO, el canal $qg$ is aun mas relevante porque el otrs canal disponible es $q\bar{q}$, cuya luminosidad está altamtente suprimida por la presencia del mar de distribuciones. En la Fig. \ref{fig:Figura10}, presentamos el cociente $R^{K/\pi}(d\sigma)/R^{K/\pi}(d_u)$ para los escenarios (1)(rojo), (2)(verde), (3)(azul) y la estimación a LO bajo el escenario (3)(negro). Si unicamente el canal $ug$ contribuye a LO, entonces el cociente $R^{K/\pi}(d\sigma)/R^{K/\pi}(d_u)$ debería ser a 1, en concordancia con Eq.(\ref{Rkpi}), la contribución total a LO tiende a 1 para valores altos de $z$, a diferencia de las desviaciones mostradas en los valores mas bajos de $z$. Es sumamente importante resaltar que las correcciones a NLO QCD al $R^{K/\pi}(d\sigma)/R^{K/\pi}(d_u)$ están también cercanos a 1, mostrando un comportamiento casi plano en el rango $z\in(0.5,0.8)$. Lo anterior nuevamente implica que las FFs para kaones podría ser directamente obtenido a través de un factor de reescalamiento de las FFs del pión, las cuales son determinadas con mayor precisión con los experimentos.
Finally, we compare the three proposed scenarios with each other, as well as with the LO estimate in scenario three. As previously mentioned, the $qg$-channel dominates photon-hadron production in colliders, with the $qQ$-channel being almost an order of magnitude smaller. 
{At LO, the $qg$-channel is the most significant contribution. Although the $q\bar{q}$-channel is enhanced in $p\bar{p}$ collisions due to valence antiquarks, it remains secondary to the $qg$ subprocess in our kinematic domain due to the dominant gluon flux and the favorable $qg \to \gamma q$ matrix elements.}
%At LO, the $qg$-channel becomes even more significant, as the only other available channel is the $q\bar{q}$-channel, whose luminosity is greatly reduced by the sea distributions. 
Fig. \ref{fig:Figura10} shows the ratio $R^{K/\pi}(d\sigma)/R^{K/\pi}(d_u)$ for {\it Scenario 1} (red), {\it Scenario 2} (green) and {\it Scenario 3} (blue), as well as the LO estimate for {\it Scenario 3} (black). If the $ug$-channel is the only one contributing at LO, the ratio $R^{K/\pi}(d\sigma)/R^{K/\pi}(d_u)$ should be 1. This agrees with Eq. (\ref{Rkpi}) and indicates that the total contribution at LO should approach 1 for high values of $z$, unlike the deviations observed at lower values of $z$.
It is important to note that NLO QCD corrections to the ratio also approach 1, exhibiting relatively constant behaviour within the $z$ range of $(0.5, 0.8)$. This again implies that the FFs for kaons could be obtained directly via a rescaling of the pion FFs, which are determined with greater precision in experiments.

{
\subsection{Subleading terms in the approximation}
\label{ssec:IIIC}
To evaluate the effectiveness of the approximation, we examine the impact of including an additional term. As previously discussed in this work, the highest luminosity originates from the $qg$ channel, with an up-quark in the final state. Consequently, the next most statistically significant contribution is that which produces a down-quark in the final state. According to Eq.~(\ref{eq:amplitudes}), the matrix elements are related such that the ratio yields:
\begin{equation}
\label{Rkd}
R^{K/\pi}(d_u,d_d)\approx \frac{4d^K_u(z_\text{REC})+d^K_d(z_\text{REC})}{4d^\pi_u(z_\text{REC})+d^\pi_d(z_\text{REC})} \, .
\end{equation}
This framework provides an estimation of the contribution from the subleading term, $d_d^h$. Figure \ref{fig:R3} displays the ratio $R^{K/\pi}(d_u,d_d)$.
\begin{figure}[h!]
    \centering
    \includegraphics[width=1\linewidth]{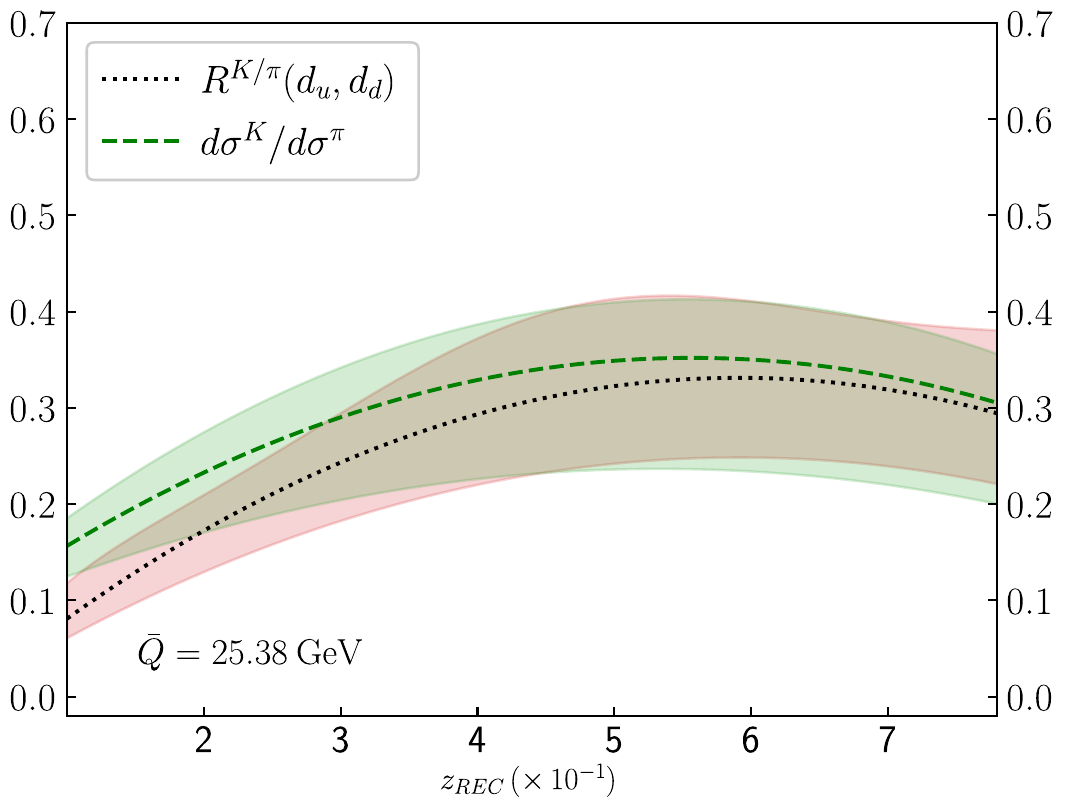}
    \caption{Same as the bottom pannel of Fig. \ref{fig:Figura7}, but for the approximated ratio $R^{K/\pi}(d_u, d_d)$, which incorporates the subleading contribution to $R^{K/\pi}(d_u)$.
    }
    \label{fig:R3}
\end{figure}
}

{
Fig. \ref{fig:R3} demonstrates that the inclusion of the subleading correction leads to improved agreement with the MC results. A closer inspection reveals that this agreement is primarily achieved when accounting for theoretical uncertainties, specifically the scale variation. While the central value of the ratio $R^{K/\pi}(d_u,d_d)$ at $\mu=p_T^\gamma$ deviates from the MC prediction, the observable must be interpreted within the context of its associated uncertainties. In this regard, the uncertainty bands provide a better agreement across the entire kinematic domain compared to the leading-order approximation, with deviations ranging between 1\% and 7\%. This result reinforces the premise that a judicious choice of kinematic cuts can yield significant information regarding FFs.}

{Finally, it should be noted that while the subleading approximation improves the agreement with MC generators, applying this methodology to claim high-precision experimental extraction would be premature. As higher-order corrections are incorporated into MC simulations and experimental measurements achieve greater precision, this analysis may necessitate more sophisticated classification techniques or the identification of alternative observables.
}
%%%%%%%%%%%%%%%%%%%%%
%%%%%%%%%%%%%%%%%%
%%%%%%%%%%%%%%%%%%%
%%%%%%%%%%%%%%%%%%%%%%%%%%%%%%%%%%%%%%%%%%
%%%%%%%%%%%%%%%%%%%%%%%%%%%%%%%%%%%%%%%%%%
%%%%%%%%%%%%%%%%%%%%%%%%%%%%%%%%%%%%
%%%%%%%%%%%%%%%%%%%%%%%%%%%%%%%%%%%%
%%%%%%%%%%%%%%%%
%%%%%%%%%%%%%%%%%%%
\section{Conclusions and outlook}
\label{sec:conclusions}

In this article, we analyzed strategies for imposing constraints on heavy meson Fragmentation Functions (FFs), using our knowledge of what happens in hadron-photon production in proton-antiproton collisions. We rely on the fact that the partonic momentum fractions can be precisely described in terms of functions that depend on quantities that can be measured experimentally, as shown in references \cite{deFlorian:2010vy,Renteria-Estrada:2022scipost}. Thus, we focused on the study of the associated $z_\text{REC}$ spectrum of $h+\gamma$ production in $p\bar{p}$ collider machines, including up to NLO QCD + LO QED effects. We considered the cases $h={\pi,K}$ since they are the lightest mesons, which involve large cross-sections and small statistical errors. We studied the ratio of production rates for $\gamma+K$ with respect to $\gamma+\pi$, as a function of $z_\text{REC}$, varying the hadron charge and other kinematic cuts. We concluded that the production of positive and negative hadrons have different behaviors when NLO QCD corrections are taken into account.

Also, we analyzed the contributions of different partonic channels, identifying the $qg$ channel as the most contributing one. With everything previously found, we proceeded to study the relation between the kaon and pion fragmentation functions and the cross-section rates, both as functions of $z_\text{REC}$. Firstly, we realized that the $ug$ channel is favored by the luminosity as well as the electromagnetic charge of $u$ quarks with respect to $d$ quarks. Then, we defined three scenarios with different kinematic configurations to subsequently analyze the relations between the ratios $R^{K/\pi}(d\sigma)=d\sigma^K/d\sigma^\pi(z_\text{REC})$ and $R^{K/\pi}(d_u)=d_u^K(z_\text{REC})/d_u^\pi(z_\text{REC})$. By imposing a pseudorapidity cut to restrict events close to the Born-type kinematics and imposing that $0.03\leq\{(x_1)_\text{REC},(x_2)_\text{REC}\}\leq 0.5$, we were able to achieve an improvement in highlighting the contribution of the $ug$ channel and isolating the $u$-initiated FFs. In fact, the results obtained in Fig. \ref{fig:Figura9} indicate that the FF ratio is strongly constrained by the corresponding cross-sections. That is, it is possible to relate the pion and kaon FFs through the computation of $R^{K/\pi}(d\sigma)$. 
{
Furthermore, we analyzed the impact of incorporating a subleading term originating from the down-quark hadronization. It was found that the overall agreement with the MC simulation is improved, thereby demonstrating the applicability of the proposed method, with observed deviations of up to 7\%.}

{Finally, this work complements previous studies conducted in the context of $pp$ collisions. In particular, we exploit the asymmetric configuration of the Tevatron ($p\bar{p}$), which dictates that the trends for the NLO QCD + LO QED cross-section ratios between kaons and pions remain monotonically increasing or nearly flat, regardless of the meson's electric charge. Furthermore, this study demonstrates that although the center-of-mass energy at the Tevatron is an order of magnitude lower than that of the LHC, this lower energy scale does not preclude the identification of strong correlations between Fragmentation Functions. Given that the electromagnetic interaction is enhanced in this specific reaction channel, the information extracted is as significant as that obtained from $pp$ collisions.}
%Finally, we conclude that the strategy used in this work constitutes a first step for the extraction of partonic channels by imposing kinematic cuts in $p\bar{p}$ collisions.

%%%%%%%%%%%%%%%%%%%%%%%%%%%%%%%%%%%%%%%%%%%%%%%%%%%%%%%%%%%%%%%%%%%%%%%%%%%%%%%%%
%%%%%%%%%%%%%%%%%%%%%%%%%%%%%%%%%%%%%%%%%%%%%%%%%%%%%%%%%%%%%%%%%%%%%%%%%%%%%%%%%
\section*{Acknowledgements}
{The authors are grateful to the reviewer for their insightful comments and suggestions, which have significantly improved the quality of the manuscript.} R.~J.~Hernández-Pinto thanks SECIHTI for the support received through Project CBF2023-2024-268 from the 2023-2024 frontier science call and by Sistema Nacional de Investigadores.  M. A. P\'erez de Le\'on is supported by SECIHTI (M\'exico) through the {\it{Estancias Posdoctorales por M\'exico}} program. D. F. Renter\'ia Estrada. is supported by Generalitat Valenciana CIGRIS/2022/145.

%%%%%%%%%%%%%%%%%%%%%%%%%%%%%%%%%%%%%%%%%%%%%%%%%%%%%%%%%%%%%%%%%%%%%%%%%%%%%%%%%
\bibliography{refs}
%%%%%%%%%%%%%%%%%%%%%%%%%%%%%%%%%%%%%%%%%%%%%%%%%%%%%%%%%%%%%%%%%
%%%%%%%%%%%%%%%%%%%%%%%%%%%%%%%%%%%%%%%%%%%%%%%%%%%%%%%%%%%%%%%%%
\end{document}